\documentclass[prodmode,acmtoms]{acmsmall}
\acmVolume{V}
\acmNumber{N}
\acmArticle{A}
\acmYear{YYYY}
\acmMonth{0}
\usepackage{url}
\usepackage{amssymb}
\usepackage{amsmath}
\usepackage{algorithm}
\usepackage{algpseudocode}
\usepackage{array}
\usepackage{hyperref}
\usepackage{booktabs}
\usepackage{a4wide}
\usepackage{stmaryrd}
\usepackage[numbers,square,sort&compress]{natbib}
\usepackage{graphicx}
\usepackage{verbatim}
\usepackage{anslistings}
\usepackage{rotating}
\usepackage{multicol}
\usepackage{multirow}
\usepackage{sidecap}

\usepackage{commath}
\DeclareMathOperator{\Div}{div}
\DeclareMathOperator{\Curl}{curl}
\DeclareMathOperator{\Grad}{grad}

\newcommand{\uflc}[1]{\texttt{#1}}

\newcommand{\jump}[1]{\llbracket #1 \rrbracket}
\newcommand{\avg}[1]{\langle #1 \rangle}
\newcommand{\operator}[1]{\mathrm{#1}\, }
\newcommand{\R}{\mathbb{R}}

\newcommand{\nedelec}{N\'ed\'elec}
\newcommand{\Gateaux}{G\^ateaux}
\newcommand{\foralls}{\forall\,}
\newcommand{\dx}{\dif{}x}
\newcommand{\ds}{\dif{}s}

\newcommand{\GammaD}{\Gamma_{\mathrm{D}}}
\newcommand{\GammaN}{\Gamma_{\mathrm{N}}}
\newcommand{\type}{\operatorname{type}}

\newcommand{\suchthat}{\quad | \quad}

\let\ifbolditem\iffalse
\def\bolditemtrue{\global\let\ifbolditem\iftrue}
\def\bolditemfalse{\global\let\ifbolditem\iffalse}

\newenvironment{benumerate}
  {\def\bolditem{\bolditemtrue\item}%
   \enumerate
   }
  {\endenumerate}

\title{Unified Form Language: A domain-specific language for weak
formulations of partial differential equations}
\author{MARTIN S. ALN{\AE}S \affil{Simula Research Laboratory}
        ANDERS LOGG \affil{Simula Research Laboratory and University of Oslo}
        KRISTIAN B. {\O}LGAARD \affil{Aalborg University}
        MARIE E. ROGNES \affil{Simula Research Laboratory}
        GARTH N. WELLS \affil{University of Cambridge}}
\begin{abstract}
We present the Unified Form Language (UFL), which is a domain-specific
language for representing weak formulations of partial differential
equations with a view to numerical approximation.  Features of UFL
include support for variational forms and functionals, automatic
differentiation of forms and expressions, arbitrary function space
hierarchies for multi-field problems, general differential operators
and flexible tensor algebra. With these features, UFL has been used to
effortlessly express finite element methods for complex systems of
partial differential equations in near-mathematical notation,
resulting in compact, intuitive and readable programs. We present in
this work the language and its construction. An implementation of UFL
is freely available as an open-source software library. The library
generates abstract syntax tree representations of variational
problems, which are used by other software libraries to generate
concrete low-level implementations. Some application examples are
presented and libraries that support UFL are highlighted.
\end{abstract}

\category{D.2.1}{Software Engineering}{Requirements/Specifications}[Languages]
\category{D.2.13}{Software Engineering}{Reusable Software}[Domain engineering]
\category{D.3.2}{Programming Languages}{Language Classifications}[Very high-level languages]
\category{I.1.1}{Symbolic and Algebraic Manipulation}{Expressions and Their Representation}[Representations (general and polynomial)]
\category{G.1.8}{Partial Differential Equations}{Finite element methods}
\category{G.1.4}{Quadrature and Numerical Differentiation}{Automatic differentiation}
\terms{Algorithms, Design, Languages}
\keywords{
AD,
algorithmic differentiation,
automatic functional differentiation,
discretization,
domain specific language,
DSEL,
DSL,
einstein notation,
embedded language,
FEM,
finite element method,
functional,
implicit summation,
index notation,
mixed element,
partial differential equation,
PDE,
symbolic differentiation,
tensor algebra,
weak form,
weak formulation,
}
\acmformat{}
\begin{document}
\begin{bottomstuff}
This research is partially supported by Research Council of Norway
through grant no.~209951 and a Center of Excellence grant awarded to
the Center for Biomedical Computing at Simula Research Laboratory.
Author's addresses:
M. S. Aln{\ae}s ({\tt martinal@simula.no}) and
A. Logg ({\tt logg@simula.no}) and
M. E. Rognes ({\tt meg@simula.no}),
Center for Biomedical Computing, Simula Research Laboratory,
P.O.~Box 134, 1325 Lysaker, Norway;
K. B. {\O}lgaard ({\tt kbo@civil.aau.dk}),
Department of Civil Engineering, Aalborg University,
Niels Bohrs Vej 8, 6700 Esbjerg, Denmark;
G. N. Wells ({\tt gnw20@cam.ac.uk})
Department of Engineering, University of Cambridge,
Trumpington Street, Cambridge CB2 1PZ, United Kingdom.
\end{bottomstuff}
\maketitle

\section{Introduction}
\label{sec:introduction}

We present a language for expressing variational forms of partial
differential equations (PDEs) in near-mathematical notation.  The
language, known as the Unified Form Language (UFL), inherits the
typical mathematical operations that are performed on variational
forms, thereby permitting compact and expressive computer input of
mathematical problems. The complexity of the input syntax is
comparable to the complexity of the classical mathematical
presentation of the problem. The language is expressive in the sense
that it provides basic, abstract building blocks which can be used to
construct representations of complicated problems; it offers a mostly
dimension-independent interface for defining differential equations;
and it can be used to define problems that involve an arbitrary number
of coupled fields. The language is developed with finite element
methods in mind, but most of the design is not restricted to a
specific numerical method.

UFL is a \emph{language} for expressing variational statements of
partial differential equations and does not provide a problem solving
environment. Instead, it generates abstract representations of
problems that can be used by \emph{form compilers} to create concrete
code implementations in general programming languages. There exist a
number of form compilers that generate low-level code from UFL. These
include the FEniCS Form Compiler
(FFC)~\citep{kirby:2006,RognesKirbyEtAl2009a,oelgaard:2010,logg:2012},
the SyFi Form Compiler (SFC)~\citep{alnaes:2010,alnaes:2012} and the
Manycore Form Compiler~\citep{markall:2010,MarkallEtAl2012}. From a
common UFL input, these compilers differ in the strategies used to
create and optimize a low-level implementation, and in the target
low-level language. The code generated by these form compilers can be
used in a problem solving environment, linked at compile time or
dynamically at runtime.

An example of a problem solving environment that uses code generated from
UFL input is DOLFIN~\citep{logg:2010,logg:2012b}, which is developed
as part of the FEniCS Project~\citep{fenics:book}. Users of DOLFIN
may describe a finite element discretization of a partial differential
equation in UFL, and call a form compiler such as FFC or SFC to generate
low-level code. In the case of FFC and SFC, this low-level code conforms
to the UFC specification~\citep{AlnaesLoggEtAl2009a,AlnaesLoggEtAl2012a},
which is a C++ interface for functionality related to evaluation of local
stiffness matrices, finite element basis functions and local-to-global
mappings of degrees of freedom. The UFC code may then be used by DOLFIN
to assemble and solve the linear or nonlinear system corresponding to
the finite element discretization described in UFL.

UFL is implemented as a domain-specific embedded language (DSEL) in
Python. The distinction between a DSEL and a high-level software
component lies in the level of expressiveness; UFL expressions can be
composed and combined in arbitrary ways within the language design
limits.
Paraphrasing P. Hudak~\citep{hudak1996building}, a DSEL is the ultimate
abstraction, allowing the user to reason about the program within the
domain semantics, rather than within the semantics of the
programming language.
As an embedded language, UFL relies on the parser and grammar of the
host language, Python. While it would be possible to select a
subset of the Python grammar and write a UFL parser for that
subset, we make no such restrictions in practice.
UFL is implemented as a Python module which defines
types (classes) and operators that together form an
expressive language for representing weak formulations of partial
differential equations. In addition, UFL provides a collection of
algorithms for operating on UFL expressions. By implementing UFL as a
DSEL in Python, we sacrifice some control over the syntax, but believe
that this is overwhelmingly outweighed by the advantages. First,
parsing is inherited and users may rely on all features of the Python
programming language when writing UFL code, for example to define new
operators. Second, it also permits the seamless integration of UFL
into Python-based problem solving environments. The Python interface
of the library DOLFIN is an example of this. In particular, the use of
just-in-time (JIT) compilation facilitates the incorporation of UFL in
a scripted environment without compromising the performance of a
compiled language. This is discussed in detail in~\citet{logg:2010}.

There have been a number of efforts to create domain-specific
languages for scientific computing applications. Examples include
SPL~\citep{xiong:2001} for signal processing and the Tensor
Contraction Engine~\citep{baumgartner:2005} for quantum chemistry
applications.  In the context of partial differential equations, there
have been a number of efforts to combine symbolic computing, code
generation and numerical methods. In some cases the code generation is
explicit, while in other cases, such as when employing templates,
implicit. Early examples include FINGER~\citep{wang:1986}, the
Symbolic Mechanics System~\citep{korelc:1997}, and
Archimedes~\citep{ShewchukGhattas1993}. Analysa~\citep{bagheri:2004}
is an abstract finite element framework of limited scope built upon
Scheme.  Feel++~\citep{prudhomme:www,prudhomme:2006} uses C++
templates to create an embedded language for solving partial
differential equations using finite element methods. Another example
of a domain-specific language embedded in C++ is
Sundance~\citep{long:2010}. Sundance relies heavily on automatic
differentiation to provide a problem solving environment targeted at
PDE-constrained optimization. UFL also provides automated
differentiation of functionals and variational forms, but the approach
differs in some respects from Sundance. This is discussed later in
this work.  UFL is distinguished from the aforementioned efforts by
its combination of a high level of expressiveness,
mathematically-driven abstractions, extensibility, breadth of
supported mathematical operations and embedding in a modern,
widely-used and freely available language (Python). Moreover, it is
deliberately decoupled from a code generator and problem solving
environment. This provides modularity and scope to pursue different
code generation and/or solution strategies from a common description
of a variational problem. This is highlighted by the existence of the
different form compilers that support UFL, with each targeting a
specific code generation strategy or architecture. Unlike some of the
efforts listed above, UFL is freely available under a GNU public
license (LGPLv3+).

The syntax used in UFL has its roots in FFC which was first released
in 2005. At the time, FFC filled the roles of both form language and
form compiler for the FEniCS Project. Much of the UFL syntax is
inherited from early versions of FFC, but has since been
re-implemented, generalized and extended to provide a more consistent
mathematical environment, to cover a richer class of nonlinear forms
and to provide a range of abstract algorithms, including
differentiation. FFC no longer provides an input syntax, rather it
generates code from a UFL representation. The UFL form language was
first released in 2009~\citep{Alnaes2009} and has since then been
tested on a wide range of applications. A rich and varied selection of
applications that use UFL are presented in~\citet{fenics:book}.

The remainder of this work is structured as
follows. Section~\ref{sec:concepts} summarizes the main mathematical
concepts on which UFL is based. A detailed presentation of the UFL
language is then given in Section~\ref{sec:language}. This is followed
in Section~\ref{sec:examples} by a number of examples that demonstrate
the use of UFL for a variety of partial differential equations. The
subsequent sections focus on the technical aspects of the UFL design. In
Sections~\ref{sec:representation} and~\ref{sec:algorithms}, we describe
the internal representation of UFL expressions and provide an overview
of the algorithms provided by UFL, respectively. Particular emphasis
is placed on differentiation.  Section~\ref{sec:validation} provides a
brief discussion of validation and code correctness. Some conclusions
are then drawn in Section~\ref{sec:conclusions}.

The implementation of UFL is available at
\url{https://launchpad.net/ufl}. The examples presented
in this work, including the UFL code used, are archived at
\url{http://www.dspace.cam.ac.uk/handle/1810/243981}.

\section{Mathematical concepts and scope}
\label{sec:concepts}

To clarify the notation, conventions, scope and assumptions of UFL and
this paper, we begin by defining some key concepts in mathematical
terms. We assume familiarity with variational formulations of PDEs and
finite element methods.
These variational formulations are assumed to be expressed as sums of
integrals over geometric domains. Each integrand is an expression
composed from a set of valid functions and geometric quantities, with
various operators applied. Each such function is an element of a
function space, typically, but not necessarily, a finite element
space, while the set of permitted operators include differential
operators and operators from tensor algebra. The central mathematical
abstractions, including multi-linear variational forms, tensor algebra
conventions and the finite element construction, are formally
introduced in the subsections below.

When enumerating $n$ objects, we count from $1$ to $n$, inclusive, in
the mathematical notation, while we count from $0$ to $n-1$,
inclusive, in computer code.

\subsection{Variational forms}
\label{sec:variationalforms}

UFL is centered around expressing finite element variational forms,
and in particular real-valued \emph{multi-linear forms}. A real-valued
multi-linear form $a$ is a map from the product of a given sequence
$\{V_j\}_{j=1}^{\rho}$ of function spaces:
\begin{equation}
  a : V_{\rho} \times \cdots \times V_2 \times V_1 \rightarrow \R,
\end{equation}
that is linear in each argument. The spaces $V_j$ are labeled
\emph{argument spaces}. For the case $\rho \le 2$, $V_1$ is referred
to as the \emph{test} space and $V_2$ as the \emph{trial} space.  The
\emph{arity} of a form $\rho$ is the number of argument spaces. Forms
with arity $\rho= 0$, $1$, or $2$ are named \emph{functionals},
\emph{linear forms} and \emph{bilinear forms}, respectively. Such
forms can be assembled on a finite element mesh to produce a scalar, a
vector and a matrix, respectively.  Note that the \emph{argument
  functions} $\langle v_j \rangle_{j=\rho}^1$ are enumerated backwards
such that their numbering matches the corresponding axis in the
assembled tensor.

If the form $a$ is parametrized over one or more \emph{coefficient
  functions}, we express the form as the mapping from a product of a
sequence $\{W_k\}_{k=1}^{n}$ of \emph{coefficient spaces} and the
argument spaces:
\begin{equation}
  \label{eq:generalform}
  \begin{split}
    a &: W_1 \times W_2 \times \cdots \times W_n \, \times \,
         V_{\rho} \times \cdots \times V_2 \times V_1 \rightarrow \R, \\
    a &\mapsto a(w_1, w_2, \ldots, w_n; v_{\rho}, \ldots, v_2, v_1).
  \end{split}
\end{equation}
Note that $a$ is assumed to be (possibly) non-linear in the
coefficient functions $w_k$ and linear in the argument
functions~$v_j$. For a detailed exposition on finite element
variational forms and assembly, we refer to~\citep{KirbyLogg2012} and
references therein. To make matters concrete, we here list examples of
some forms with different arity~$\rho$ and number of coefficients~$n$:
\begin{align}
  a(u, v) &:= \int_{\Omega} \Grad u \cdot \Grad v \dx,
    & \rho = 2, \quad& n = 0, \\
  a(\epsilon; u, v) &:= \int_{\Omega} \epsilon^2 \Grad u \cdot \Grad v \dx,
    & \rho = 2, \quad& n = 1, \\
  a(f; v) &:= \int_{\Omega} f v \dx,
    & \rho = 1, \quad& n = 1, \\
  a(u, v;) &:= \int_{\Omega} |\Grad (u - v)|^2 \dx
    & \rho = 0, \quad& n = 2,
\end{align}
where $\Omega$ is the geometric domain of interest.

\subsubsection{Geometric domains and integrals}
\label{subsubsec:geometric_domains_and_integrals}

UFL supports multi-linear forms defined via integration over geometric
domains in the following manner. Let $\Omega \subset \R^d$ be a domain
with boundary $\partial \Omega$ and let $\mathcal{T}= \{T\}$ be
a suitable tessellation such that $\Omega = \bigcup_{T \in \mathcal{T}}
T$. We denote the induced tessellation of $\partial \Omega$ by
$\mathcal{F} = \{F\}$, and let $\mathcal{F}^0$ denote the set of
internal facets of $\mathcal{T}$. Each of the three sets, the
\emph{cells} in $\mathcal{T}$, the \emph{exterior facets} in
$\mathcal{F}$ and the \emph{interior facets} in $\mathcal{F}^0$, is
assumed to be partitioned into one or more disjoint subsets:
\begin{equation}
  \mathcal{T} = \bigcup_{k=1}^{n_{c}} \mathcal{T}_{k},
  \quad
  \mathcal{F} = \bigcup_{k=1}^{n_{f}} \mathcal{F}_{k},
  \quad
  \mathcal{F}^0 = \bigcup_{k=1}^{n_{f}^{0}} \mathcal{F}_{k}^0,
\end{equation}
where $n_c$, $n_f$ and $n_{f}^0$ denote the number of subsets of
cells, exterior facets and interior facets, respectively. Given these
definitions, it is assumed that the multi-linear form can be expressed
in the following canonical form:
\begin{equation}
  \label{eq:canonical}
  \begin{split}
    a(w_1, w_2, \ldots, w_n; v_{\rho}, \ldots, v_2, v_1)
    &=
    \sum_{k=1}^{n_c} \sum_{T\in\mathcal{T}_{k}} \int_{T}
    I^c_k(w_1, w_2, \ldots, w_n; v_{\rho}, \ldots, v_2, v_1) \dx
    \\
    &+
    \sum_{k=1}^{n_f} \sum_{F\in\mathcal{F}_{k}} \int_{F}
    I^f_k(w_1, w_2, \ldots, w_n; v_{\rho}, \ldots, v_2, v_1) \ds
    \\
    &+
    \sum_{k=1}^{n_f^0} \sum_{F\in\mathcal{F}_{k}^0} \int_{F}
    I^{f,0}_k(w_1, w_2, \ldots, w_n; v_{\rho}, \ldots, v_2, v_1) \ds,
  \end{split}
\end{equation}
where $\dx$ and $\ds$ denote appropriate measures. The integrand
$I^{c}_{k}$ is integrated over the $k$th subset $\mathcal{T}_{k}$ of
cells, the integrand $I^{f}_{k}$ is integrated over the $k$th subset
$\mathcal{F}_{k}$ of exterior facets and the integrand $I^{f, 0}_{k}$
is integrated over the $k$th subset $\mathcal{F}_{k}^0$ of interior
facets.

UFL in its current form does not model geometrical domains, but allows
integrals to be defined over subdomains associated with an integer
index~$k$. It is then the task of the user, or the problem-solving
environment, to associate the integral defined over the subdomain $k$
with a concrete representation of the geometrical subdomain.

\subsubsection{Differentiation of forms}

Differentiation of variational forms is useful in a number of
contexts, such as the formulation of minimization problems and
computing Jacobians of nonlinear forms. In UFL, the derivative of a
form is based on the \Gateaux{} derivative as detailed below.

Let $f$ and $v$ be coefficient and argument functions, respectively,
with compatible domain and range. Considering a functional $M = M(f)$, the
\Gateaux{} derivative of $M$ with respect to $f$ in the direction $v$ is
defined by
\begin{equation}
  \label{eq:diffdef}
  M^{\prime}(f; v) \equiv D_f M(f)[v] = \frac{d}{d\tau} \left[M(f + \tau v)\right]_{\tau=0}.
\end{equation}
Given a linear form $L(f; v)$ (which could be the result of the above
derivation) and another compatible argument function $u$, we can
continue by computing the bilinear form $L^{\prime}(f; u, v)$; that
is, the derivative of $L$ with respect to $f$ in the direction $u$,
defined by
\begin{align}
  L^{\prime}(f; u, v) \equiv
  D_f L(f; v)[u]
  = \frac{d}{d\tau} \left[L(f + \tau u; v)\right]_{\tau=0}.
\end{align}
In general, this process can be applied to forms of general arity $\rho
\geq 0$ to produce forms of arity~$\rho + 1$. Note that if the form to
be differentiated involves an integral, we assume that the integration
domain does not depend on the differentiation variable. To express
the differentiation of a general form, consider the following compact
representation of the canonical form~\eqref{eq:canonical}:
\begin{align}
  \label{eq:generalform2}
  F(\langle w_i \rangle_{i=1}^{n}; \langle v_j \rangle_{j=\rho}^{1})
  = \sum_k \int_{D_k} I_k(\langle w_i \rangle_{i=1}^{n};
  \langle v_j \rangle_{j=\rho}^{1}) \dif \mu_k,
\end{align}
where $\left\{D_k\right\}$ and $\left\{\dif\mu_k\right\}$ are the
geometric domains and corresponding integration measures, and
$\left\{I_k\right\}$ are the integrand expressions. We can then write
the derivative of the general form~\eqref{eq:generalform2} with
respect to, for instance, $w_1$ in the direction $v_{\rho+1}$ as
\begin{align}
  \label{eq:generalderivative}
D_{w_1}F(\langle w_i \rangle_{i=1}^{n}; \langle v_j \rangle_{j=\rho}^{1})[v_{\rho+1} ]
    = \sum_k \int_{D_k} \frac{d}{d\tau} \left[
      I_k(w_1 + \tau v_{\rho+1}, \langle w_i \rangle_{i=2}^{n};
      \langle v_j \rangle_{j=\rho}^{1}) \right]_{\tau=0} \, d\mu_k.
\end{align}

\subsection{Tensors and tensor algebra}

A core feature of UFL is its tensor algebra support. We summarize here
some elementary tensor algebra definitions, notation and operations that
will be used throughout this paper.

First, an \emph{index} is either a fixed positive\footnote{Indices are
  positive in the mathematical base 1 notation used here and
  non-negative in the base 0 notation used in the computer code.}
integer value, in which case it is labeled a \emph{fixed-index}, or a
symbolic index with no value assigned, in which case it is called a
\emph{free-index}. A \emph{multi-index} is an ordered tuple of
indices, each of which can be free or fixed. Moreover, a
\emph{dimension} is a strictly positive integer value. A \emph{shape} $s$ is an
ordered tuple of zero or more dimensions: $s = (s_1, \dots, s_r)$; the
corresponding rank $r \geqslant 0$ equals the length of the shape
tuple.

Any tensor can be represented either as a mathematical object with a
(tensor) shape or in terms of its scalar
components with reference to a given basis. More precisely, the
following notation and bases are used. Scalars are considered rank
zero tensors. We denote by $\{ e^i \}_{i=1}^{d}$ the standard
orthonormal Euclidean basis for $\R^d$ of dimension $d$. A basis
$\{E^{\alpha}\}_{\alpha}$ for $\R^{s}$, where $s = (s_1, \dots, s_r)$
is a shape, is naturally defined via outer products of the vector
basis:
\begin{equation}
  E^{\alpha} \equiv e^{\alpha_1} \otimes \cdots \otimes e^{\alpha_r},
\end{equation}
where the range of the multi-index $\alpha = (\alpha_1, \dots,
\alpha_r)$ is such that $1 \leqslant \alpha_i \leqslant s_i$ for $i =
1, \dots, r$. In general, whenever a multi-index $\alpha$ is used to
index a tensor of shape $s$, it is assumed that $1 \leqslant
\alpha_i \leqslant s_i$ for $i = 1, \dots, r$. Then, the scalar
component of index $i$ of a vector $v$ defined relative to the basis
$\{e^i\}_i$ is denoted $v_i$. More generally, for a tensor $C$ of
shape $s$, $C_{\alpha}$ denotes its scalar component of multi-index
$\alpha$ with respect to the basis $\{ E^{\alpha} \}_{\alpha}$ of
$\R^s$. Moreover, whenever we write $\sum_i v_i$, we imply
$\sum_{i=1}^{d} v_i$, where $d$ is the dimension of
$v$. Correspondingly, for sums over multi-indices, $\sum_\alpha
C_\alpha$ implies $\sum_{\alpha_1}\cdots\sum_{\alpha_r} C_\alpha$ with
the deduced ranges.

Whenever one or more free-indices appear twice in a monomial term,
summation over the free-indices is implied. Tensors $v$, $A$ and $C$
of rank 1, 2 and $r$, respectively, can be expressed using the summation
convention as:
\begin{equation}
  v = v_i e^i,
  \quad
  A = A_{ij} E^{ij},
  \quad
  C = C_{\alpha} E^{\alpha}.
\end{equation}

We will also consider general tensor-valued functions $f: \Omega
\rightarrow \R^s$, $\Omega \subset \R^d$, where the shape $s$ in this
context is termed the \emph{value shape}. Indexing of tensor-valued
functions follows the same notation and assumptions as for
tensors. Furthermore, derivatives with respect to spatial coordinates
may be compactly expressed in index notation with the comma convention
in subscripts. For example, for coordinates $(x_1, \dots, x_d) \in
\Omega$ and a function $f: \Omega \rightarrow \R$, a vector function
$v: \Omega \rightarrow \R^{n}$ or a tensor function $C: \Omega
\rightarrow \R^{s}$, we write for indices $i$ and $j$, and multi-indices
$\alpha$ and $\beta$, with the length of $\beta$ denoted by~$p$:
\begin{equation}
  f_{,i} \equiv \frac{\partial f}{\partial x_i},
  \quad
  v_{i,j} \equiv \frac{\partial v_i}{\partial x_j},
  \quad
  C_{\alpha, \beta} \equiv
    \frac{\partial^p C_\alpha}{\partial x_{\beta_1} \cdots
          \partial x_{\beta_{p}}}.
\end{equation}
\subsection{Finite element functions and spaces}
\label{sec:finite_element}

A finite element space $V_h$ is a linear space of piecewise
polynomial fields defined relative to a tessellation $\mathcal{T}_h =
\{ T \}$ of a domain $\Omega$. Such spaces are typically defined
locally; that is, each field in the space is defined by its
restriction to each cell of the tessellation. More precisely, for a
finite element space $V_h$ of tensor-valued functions of value
shape $s$, we assume that
\begin{equation}
  \label{eq:finiteelement}
  V_h = \{v \in \mathcal{H} \,:\, v |_{T} \in \mathcal{V}_{T} \}, \quad
  \mathcal{V}_{T} = \{v : T \rightarrow \R^s \,:\, v_{\alpha} \in \mathcal{P}(T) \quad\foralls \alpha\},
\end{equation}
where the space $\mathcal{H}$ indicates the global regularity and
where $\mathcal{P} = \mathcal{P}(T)$ is a specified (sub-)space of
polynomials of degree $q \geqslant 0$ defined over~$T$. In other
words, the global finite element space $V_h$ is defined by patching
together local finite element spaces $\mathcal{V}_T$ over the
tessellation $\mathcal{T}_h$. Note that the polynomial spaces may vary
over the tessellation; however, this dependency is usually omitted for
the sake of notational brevity.

The above definition may be extended to mixed finite element spaces.
For given local finite element spaces $\{\mathcal{V}_i\}_{i=1}^{n}$ of
respective value shapes $\{s^i\}_{i=1}^{n}$, we define the mixed local
finite element space $\mathcal{W}$ by:
\begin{equation}
  \mathcal{W} = \mathcal{V}_1 \times \mathcal{V}_2 \times \cdots \times \mathcal{V}_n
  = \{w = (v_1, v_2, \ldots, v_n) \,:\, v_i \in \mathcal{V}_i, \quad i=1,2,\ldots,n\}.
\end{equation}
The extension to global mixed finite element spaces follows as
in~\eqref{eq:finiteelement}.  Note that all element factors in a mixed
element are assumed to be defined over the same cell~$T$. The
generalization to nested hierarchies of mixed finite elements follows
immediately, and hence such hierarchies are also admitted

Any $w \in \mathcal{W}$ has the representation $w: T \rightarrow
\R^{t}$, with a suitable shape tuple~$t$, and the value components of
$w$ must be mapped to components of $v_i \in \mathcal{V}_i$. Let $r^i$
be the corresponding rank of the value shape $s^i$, and denote by
$p^i=\prod_{j=1}^{r^i} s^i_j$ the corresponding \emph{value size};
i.e, the number of scalar components. In the general case, we choose a
rank~$1$ shape $t = (\sum_{i=1}^n p^i)$, and map the flattened
components of each $v^i$ to components of the vector-valued $w$; that
is,
\begin{align}
  w = (v^1_1, \ldots, v^1_{p^1}, \ldots, v^n_1, \ldots, v^n_{p^n}).
\end{align}
This mapping permits arbitrary combinations of value shapes $s^i$. In
the case where $\mathcal{V}_i$ coincide for all $i = 1, \dots, n$, we refer to
the resulting specialized mixed element as a vector element of
dimension $n$, and choose $t=(n, s^1_1, \ldots, s^1_{r^1})$. As a
generalization of vector elements, we allow tensor
elements\footnote{Tensor-valued elements, not to be confused with
  tensor product elements.} of shape $c$ with rank~$q$, which gives
the value shape $t=(c_1, \ldots, c_q, s^1_1, \ldots,
s^1_{r^1})$. Tensor elements built from scalar subelements may have
symmetries, represented by a symmetry mapping from component to
component.  The number of subelements then equals
$n=\left(\prod_{i=1}^{q} c_i\right)-m$, where $m$ is the number of
value components that are mapped to another.

For some finite element discretizations, it is helpful to represent
the local approximation space as the \emph{enrichment} of one element
with another. More precisely, for local finite element spaces
$\mathcal{V}_1, \mathcal{V}_2, \dots, \mathcal{V}_n$ defined over a
common cell $T$ and of common value shape $s$, we define the space
\begin{equation}
  \mathcal{W} = \mathcal{V}_1 + \mathcal{V}_2 + \dots + \mathcal{V}_n
  = \{ v_1 + v_2 + \dots + v_n \,:\, v_i \in V_i, \quad i = 1, \dots, n \}.
\end{equation}
Again, the extension to global enriched finite element spaces follows
as in~\eqref{eq:finiteelement}. The MINI
element~\citep{ArnoldBrezziFortin1984} for the Stokes equations is an
example of an enriched element.

\section{Overview of the language}
\label{sec:language}

UFL can be partitioned into sublanguages for finite elements, expressions,
and forms. We will address each separately below. Overall, UFL has a
declarative nature similar to functional programming languages. Side
effects, sequences of statements, subroutines and explicit loops
found in imperative programming languages are all absent from UFL. The
only branching instructions are inline conditional expressions, which
will be further detailed in Section~\ref{sec:conditionaloperators}.

\subsection{Finite elements}
\label{sec:function_spaces}

The UFL finite element sublanguage provides syntax for finite
elements and operations over finite elements, including mixed and
enriched finite elements, as established in
Section~\ref{sec:finite_element}.

\subsubsection{Finite element abstractions and classes}

UFL provides four main finite element abstractions: primitive finite
elements, mixed finite elements, enriched finite elements and
restricted finite elements. Each of these abstractions provides
information on the value shape, the cell and the embedding polynomial
degree of the element (see~\eqref{eq:finiteelement}), and each is
further detailed below. We remark that UFL is primarily concerned with
properties of local finite element spaces: the global continuity
requirement and the specific implementation of the element degrees of
freedom or basis functions are not covered by UFL. For an overview of
finite element abstractions with initialization arguments,
see~Table~\ref{tab:finiteelementclasses}. Example usage will be
presented in Section~\ref{sec:examples}.

\begin{table}
  \tbl{
  Overview of element classes defined in UFL.
}{
  \begin{tabular}{l}
    \toprule
    Finite element class specification\\
    \midrule
    \uflc{FiniteElement(family, cell, degree)} \\
    \uflc{VectorElement(family, cell, degree, (dim))} \\
    \uflc{TensorElement(family, cell, degree, (shape), (symmetry))} \\
    \midrule
    \uflc{MixedElement(elements)} \\
    \uflc{EnrichedElement(elements)} \\
    \uflc{RestrictedElement(element, domain)} \\
    \bottomrule
  \end{tabular}
}
\tabnote{In addition to
    the arguments given here, a specific quadrature scheme can be
    given for the primitive finite elements (those defined by family
    name). Arguments in parentheses are optional.}
  \label{tab:finiteelementclasses}
\end{table}

In the literature, it is common to refer to finite elements by their
family parametrized by cell type and order: for instance, the
``\nedelec{} face elements of the second kind over tetrahedra of second
order''~\citep{Nedelec1986}. The global continuity requirements are
typically implied by the family: for instance, it is
generally assumed that the aforementioned \nedelec{} face element
functions indeed do have continuous normal components across
faces. Moreover, finite elements may be known by different family
names, for instance the aforementioned \nedelec{} face elements
coincide with the Brezzi--Douglas--Marini elements on
tetrahedra~\citep{BrezziDouglasMarini1985}, which again coincide with
the $\mathcal{P} \Lambda^2(T)$ family on
tetrahedra~\citep{ArnoldFalkWinther2006}.

UFL mimics the literature in the sense that primitive finite
elements are defined in terms of a family, a cell and a
polynomial degree via the \uflc{FiniteElement} class (see
Table~\ref{tab:finiteelementclasses}). Additionally, a quadrature
scheme label can be given as an optional argument. The family must be
an identifying string, while the cell is a description of the cell
type of the geometric domain.
The UFL documentation contains the comprehensive list of
preregistered families and cells. Multiple names in the literature for
the same finite element are handled via family aliases. UFL supports
finite element exterior calculus notation for simplices in one, two or
three dimensions via such aliases. By convention, elements of a finite
element family are numbered in terms of their polynomial degree~$q$
such that their fields are indeed included in the complete polynomial
space of degree~$q$. This facilitates internal consistency, although
it might conflict with some notation in the literature. For instance,
the lowest order Raviart--Thomas elements have degree~$1$ in UFL.

Syntax is provided for defining vector elements. The \uflc{VectorElement}
class accepts a family, a cell, a degree and the dimension of the
vector element. The dimension defaults to the geometric dimension $d$
of the cell. The value shape of a vector element is then $(d, s)$
where $s$ is the value shape of the corresponding finite element
of the same family. This corresponding element may be vector-valued,
e.g. a \uflc{VectorElement("BDM", triangle, p)} has value shape
$(2,2)$. Moreover, further structure can be imposed for
(higher-dimensional) vector elements with a rank two tensor structure. The
\uflc{TensorElement} class accepts a family, a cell and a degree, and in
addition a shape and a symmetry argument. The shape argument defaults
to the tuple $(d, d)$ and the value shape of the tensor element is
then~$(d, d, s)$. The symmetry argument may be boolean true to define
the symmetry $A_{ij} = A_{ji}$ if the value rank is two. It may also
be a mapping between the component tuples that should be equal, such
as \uflc{\{(0,0):(0,1),
  (1,0):(1,1)\}} to define the symmetries $A_{11}=A_{12}$,
$A_{21}=A_{22}$.  The vector and tensor element classes can be viewed
as optimized, special cases of mixed finite elements.

In general, mixed finite elements in UFL are created from a tuple of
subelements through the \uflc{MixedElement} class. Each subelement can
be a finite, vector, or tensor element as described above, or in turn
a general mixed element. The latter can lead to nested mixed finite
elements of arbitrary, though finite, depth. All subelements must be
defined over the same geometric cell and utilize the same quadrature
scheme (if prescribed). The degree of a mixed finite element is
defined to be the maximal degree of the subelements. Note that mixed
finite elements are recursively flattened. Their value shape is $(s,
)$ where $s$ is the total number of scalar components.

Enriched elements can be defined via the \uflc{EnrichedElement} class,
given a tuple of finite, vector, tensor, or mixed subelements. The
subelements must be defined on the same cell and have the same value
shape. These then define the cell and value shape of the enriched
element. The degree is inferred as the maximal degree of the
subelements.

Finally, UFL also offers a restricted element abstraction via the
\uflc{RestrictedElement} class, taking as arguments any of the element
classes described above and a cell or the string \uflc{"facet"}.
The term restricted in this
setting refers to the elimination of element functions that vanish on
the given cell entities; \citet{labeur:2012} provide an example utilizing
elements restricted to cell facets. The value shape, cell and degree
of a restricted element are directly deduced from the defining
element.

\subsubsection{Operators over finite elements}

For readability and to reflect mathematical notation, UFL provides
some operators over the finite element classes defined in the previous
section. These operators include the binary operators
multiplication~(\uflc{*}) and addition~(\uflc{+}), and an indexing
operator (\uflc{[]}). These operators and their long-hand equivalents
are presented in Table~\ref{tab:finiteelementoperations}. The
multiplication operator acts on two elements to produce a mixed
element with the two elements as subelements in the given order. Note
that the multiplication operator (in Python) is binary, so
multiplication of three or more elements produces a nested mixed
element. Similarly, the addition operator acts on two elements to
yield an enriched element with the two given elements as
subelements. Finally, the indexing operator restricts an element to
the cell entity given by the argument to \uflc{[]}, thus returning a restricted
element.

\begin{table}
  \tbl{
    An overview of UFL operators over elements: examples of
    operator usage matched with the equivalent verbose syntax.
}{
  \begin{tabular}{ll}
    \toprule
    Operation & Equivalent syntax \\
    \midrule
    \uflc{M = U * V}     & \uflc{M = MixedElement(U, V)} \\
    \uflc{M = U * V * W} & \uflc{M = MixedElement((U, V), W)} \\
    \uflc{M = U + V}     & \uflc{M = EnrichedElement(U, V)} \\
    \uflc{M = V['facet']}& \uflc{M = RestrictedElement(V, 'facet')} \\
    \bottomrule
  \end{tabular}
}
  \label{tab:finiteelementoperations}
\end{table}

\subsection{Expressions}
\label{sec:expressions}

The language for declaring expressions consists of a set of terminal
expression types and a set of operators acting on other expressions.
Each expression is represented by an object of a subclass of \uflc{Expr}.
Each operator acts on one or more expressions and produces a new
expression. An operator result is uniquely defined by its operator
type and its operand expressions, and cannot have non-expression data
associated with it. A terminal expression does not depend on any other
expressions and typically has non-expression data associated with it,
such as a finite element, geometry data or the values of literal
constants. Terminal expression types are subclasses of \uflc{Terminal} and
operator results are represented by subclasses of \uflc{Operator}, both of
which are subclasses of \uflc{Expr}. Any UFL expression object is the root of
a self-contained expression tree in which each tree node is an \uflc{Expr}
object. The references from objects of \uflc{Operator} subtypes to the
operand expressions represent directed edges in the tree and objects
of \uflc{Terminal} subtypes terminate the tree.

As an embedded language, UFL allows the use of Python variables to
store subexpression references for reuse. However, UFL itself does not
have the concept of mutable variables. In fact, a key property of all
UFL expressions, including terminal types, is their immutable
state.\footnote{In the PyDOLFIN library, UFL function types are
  subclassed to carry additional mutable state which does
  not affect their symbolic meaning. UFL algorithms carefully
  preserve this information.}
Immutable state is a prerequisite for
the reuse of subexpression objects in expression trees by reference
instead of by copying. This aspect is critical for an efficient
symbolic software implementation.

The dependency set of an expression is the set of non-literal terminal
expressions that can be reached from the expression root.  An expression
with an empty dependency set can be evaluated symbolically, but in
general the evaluation of a UFL expression can only be carried out when
the values of its dependencies are known. Numerical evaluation of the
symbolic expression without code generation is possible when such values
are provided, but this is an expensive operation and not suitable for
large scale numerical computations.

Every expression is considered to be tensor-valued and its shape
must always be defined. Furthermore, every expression has a set of
free-indices.  Note that the free-index set of any particular expression
object is not associated with its shape; for instance, if $A$ is a rank
two tensor with shape $(3,3)$ (and no free indices), then $A_{ij}$ is
a rank zero tensor expression; in other words, scalar-valued and with
the associated free-indices $i$ and~$j$.  Mathematically one could
see $A$ and $A_{ij}$ as being the same, but represented as objects
in software they are distinct.  While $A$ represents a matrix-valued
expression, $A_{ij}$ represents any scalar value of~$A$. Because the
tensor properties of all subexpressions are known, dimension errors and
inconsistent use of free-indices can be detected early. The following
sections describe terminal expressions, the index notation and various
operators in more detail.

\subsubsection{Terminal expressions}

Terminal expressions in UFL include literal constants, geometric
quantities and functions. In particular, UFL provides a
domain-specific set of types within these three fairly generic groups.
Tables~\ref{tab:literalexpr} and~\ref{tab:terminalexpr} provide an overview of the literal
constants, geometric quantities and functions available.

Literal constants include integer and real-valued constants, the
identity matrix, the Levi--Civita permutation symbol and unit
tensors. Geometric quantities include spatial coordinates and cell
derived quantities, such as the facet normal, facet area, cell volume,
cell surface area and cell circumradius. Some of these are only well
defined when restricted to facets, so appropriate errors are emitted
if used elsewhere.

Functions are cell-wise or spatially varying expressions. These are
central to the flexibility of UFL. However, in contrast to other UFL
expressions, functions are merely symbols or placeholders. Their values
must generally be determined outside of UFL. All functions are defined
over function spaces, introduced in Section~\ref{sec:function_spaces},
such that their tensor properties, including their shape, can be derived
from the function space. Functions are further grouped into coefficient
functions and argument functions. Expressions must depend linearly on
any argument functions; see Section~\ref{sec:variationalforms}. No
such limitations apply to dependencies on geometric quantities or
coefficient functions. Functions are counted, or assigned a count,
as they are constructed, and so the order of construction matters. In
particular, different functions are assumed to have different counts.
The ordering of the arguments to a form of rank 2 (or higher) is
determined by the ordering of these counts.
For convenience, UFL provides constructors for argument functions called
\uflc{TestFunction} and \uflc{TrialFunction} which apply a fixed
ordering to avoid accidentally transposing bilinear forms.

\begin{table}
  \tbl{
Tables of literal tensor constants.
}{
\begin{tabular}{ll}
\toprule
Mathematical notation  & UFL notation \\
\midrule
 $I$                                 & \uflc{I = Identity(2)} \\
 $\epsilon$                          & \uflc{eps = PermutationSymbol(3)} \\
 $e_x, e_y$                           & \uflc{ex, ey = unit\_vectors(2)} \\
 $e_x e_x, e_x e_y, e_y e_x, e_y e_y$  & \uflc{exx, exy, eyx, eyy = unit\_matrices(2)} \\
\bottomrule \\
\end{tabular}
}
  \label{tab:literalexpr}
\end{table}

\begin{table}
  \tbl{
Tables of non-literal terminal expressions.
}{
\begin{tabular}{ll|ll}
\toprule
\multicolumn{2}{c}{\emph{Geometric quantities}}
& \multicolumn{2}{c}{\emph{Functions}} \\
\midrule
Math. & UFL notation & Math. & UFL notation \\
\midrule
 $x$                         & \uflc{x = cell.x} &
   $c \in \R$                & \uflc{c = Constant(cell)} \\
 $n$                         & \uflc{n = cell.n} &
   $g \in \mathcal{V}$                 & \uflc{g = Coefficient(element)} \\
 $|T|$                       & \uflc{h = cell.volume} &
   $w \in \mathcal{V}$                 & \uflc{w = Argument(element)} \\
 $r(T)$                      & \uflc{r = cell.circumradius} &
   $u \in \mathcal{V}$                 & \uflc{u = TrialFunction(element)} \\
 $|F|$                       & \uflc{fa = cell.facetarea} &
   $v \in \mathcal{V}$                 & \uflc{v = TestFunction(element)} \\
 $\sum_{F \subset T} |F|$       & \uflc{ca = cell.cellsurfacearea} &
 {} & {} \\
\bottomrule
\end{tabular}
}
\tabnote{The examples are given with
  reference to a predefined cell $T \subset \R^d$ denoted \uflc{cell},
  with coordinates $x \in \R^d$, facets $\{F\}$ and facet normal $n$,
  and a predefined local finite element space
  $\mathcal{V}$ of some finite element denoted
  \uflc{element}. $|\cdot|$ denotes the volume, while $r(T)$ denotes
  the circumradius of the cell $T$; that is, the radius of the
  circumscribed sphere of the cell.}
  \label{tab:terminalexpr}
\end{table}

In the FEniCS pipeline, functions are evaluated as part of the form
compilation or the assembly process. Argument functions are
interpreted by the form compiler as placeholders for each basis
function in their corresponding local finite element spaces, which are
looped over when computing the local element tensor.
Moreover, the ordering of the argument functions (defined by
their counts) determines which global tensor axis each argument is
associated with when assembling the global tensor (such as a sparse
matrix) from a form. Form compilers typically specialize the evaluation of
the basis functions during compilation, but may theoretically keep the
choice of element space open until runtime.
On the other hand, coefficient functions are
used to represent global constants, finite element fields or any
function that can be evaluated at spatial coordinates during finite
element assembly. The limitation to functions of spatial coordinates
is necessary for the integration of forms to be a cell-wise operation.

\subsubsection{Index notation}
\label{subsubsec:index_notation}

UFL mirrors conventional index notation by providing syntax for
defining fixed and free-indices and for indexing objects. For
convenience, free-index objects with the names
\uflc{i, j, k, l, p, q, r, s} are predefined. However, these can be
redefined and new ones created. A single free-index object is created
with \uflc{i = Index()}, while multiple indices are created with
\uflc{j, k, l = indices(3)}.

The main indexing functionality of UFL is summarized in
Table~\ref{tab:indexingoperators}. The indexing operator \uflc{[]}
applied to an expression yields a component-wise representation. For
instance, for a rank two tensor $A$ (\uflc{A}) and free-indices
\uflc{i}, \uflc{j}, \uflc{A[i, j]} yields the component-wise
representation~$A_{ij}$. The mapping from a scalar-valued expression
with components identified by free-indices to a tensor-valued
expression is performed via \uflc{as\_tensor(A[i, j], (i, j))}.  The
\uflc{as\_vector}, \uflc{as\_matrix}, \uflc{as\_tensor}
functions can also be used to construct tensor-valued expressions from
explicit tuples of scalar components. Note how the combination of
indexing and \uflc{as\_tensor} allows the reordering of the tensor
axes in the expression~$A_{ijkl} = B_{klij}$. Finally, we remark that
fixed and free-indices can be mixed during indexing and that standard
slicing notation is available.
\begin{table}
  \tbl{
Table of indexing operators:
  \uflc{i}, \uflc{j}, \uflc{k}, \uflc{l} are free-indices, while
  \uflc{a}, \uflc{b}, \uflc{c} are other expressions.
}{
\begin{tabular}{ll}
\toprule
Mathematical notation & UFL notation \\
\midrule
 $A_{i}$
& \uflc{A[i]} \\
 $B_{ijkl}$
& \uflc{B[i,j,k,l]} \\
 $\langle a, b, c \rangle$
& \uflc{as\_vector((a, b, c))} \\
 $A = B_i e_i, \quad (A_i = B_i)$
& \uflc{as\_vector(B[i], i)} \\
 $A = B_{ji} E^{ij}, \quad (A_{ij} = B_{ji})$
& \uflc{as\_matrix(B[j,i], (i,j))} \\
 $A = B_{klij} E^{ijkl}, \quad (A_{ijkl} = B_{klij})$
& \uflc{as\_tensor(B[k,l,i,j], (i,j,k,l))} \\
\bottomrule
\end{tabular}
}
 \label{tab:indexingoperators}
\end{table}

\subsubsection{Arithmetic and tensor algebraic operators}

UFL defines arithmetic operators such as addition, multiplication,
$l^2$ inner, outer and cross products and tensor algebraic operators
such as the transpose, the determinant and the inverse. An overview of
common operators of this kind is presented in
Table~\ref{tab:tensoroperators}. These operators will be familiar to
many readers and we therefore make only a few comments below.

\begin{table}
  \tbl{
Table of tensor algebraic operators
}{
\begin{tabular}{ll}
\toprule
Math. notation & UFL notation \\
\midrule
 $A + B$         & \uflc{A + B} \\
 $A \cdot B$         & \uflc{dot(A, B)} \\
 $A : B$             & \uflc{inner(A, B)} \\
 $A B \equiv A \otimes B$       & \uflc{outer(A, B)} \\
 $A \times B$        & \uflc{cross(A, B)} \\
\midrule
 $A^T$                         & \uflc{transpose(A)} \\
 $\operator{sym} A$            & \uflc{sym(A)} \\
 $\operator{skew} A$           & \uflc{skew(A)} \\
 $\operator{dev} A$            & \uflc{dev(A)} \\
 $\operator{tr} \equiv A_{ii}$  & \uflc{tr(A)} \\[0.35em]
\bottomrule
\end{tabular}
\hspace{0.5cm}
\begin{tabular}{ll}
\toprule
Math. notation & UFL notation \\
\midrule
 $\operator{det} A$            & \uflc{det(A)} \\
 $\operator{cofac} A$          & \uflc{cofac(A)} \\
 $A^{-1}$                       & \uflc{inv(A)} \\
\midrule
 $v \suchthat v_i = A_{ii} \, \text{(no sum)}$     & \uflc{diag\_vector(A)} \\
 $A \suchthat A_{ij} = \begin{cases} B_{ij}, \, \text{if } i = j,\\ 0, \, \text{otherwise} \end{cases}$
                               & \uflc{diag(B)} \\
 $A \suchthat A_{ij} = \begin{cases} v_i, \, \text{if } i = j,\\ 0, \, \text{otherwise} \end{cases}$
                               & \uflc{diag(v)} \\
\, \\
\bottomrule
\end{tabular}
}
 \label{tab:tensoroperators}
\end{table}

Addition and subtraction require that the left and right operands have
the same shape and the same set of free-indices. The result inherits
those properties.

In the context of tensor algebra, the concept of a product is heavily
overloaded. Therefore, the product operator \uflc{*} has no unique
intuitive definition. Our choice in UFL is to define \uflc{*} as the
product of two scalar-valued operands, one scalar and one tensor
valued operand, and additionally as the matrix--vector and
matrix--matrix product. The \uflc{inner} function defines the inner
product between tensors of the same shape, while \uflc{dot} acts on
two tensors by contracting the last axis of the first argument and the
first axis of the second argument. If both arguments are
vector-valued, the action of \uflc{dot} coincides with that of
\uflc{inner}.

When applying the product operator \uflc{*} to operands with free indices,
summation over repeated free-indices is implied. Implicit summation
is only allowed when at least one of the operands is scalar-valued,
and the tensor algebraic operators assume that their operands have no
repeated free-indices.

\subsubsection{Nonlinear scalar functions}

UFL provides a number of familiar nonlinear scalar real functions,
listed in Tables~\ref{tab:elementarytrigfunctions}
and~\ref{tab:specialfunctions}. All of these elementary,
trigonometric or special functions assume a scalar-valued expression
with no free-indices as argument. Their mathematical meaning is well
established and implementations are available in standard C++ or
Boost~\citep{boost:www}. To apply any scalar function to the
components of a tensor-valued expression, the element-wise operators
listed in Table~\ref{tab:specialfunctions} can be used.
\begin{table}
\tbl{
(Left) Table of elementary, nonlinear functions.
(Right) Table of trigonometric functions.
}{
  \begin{tabular}{ll}
    \toprule
    Math. notation & UFL notation \\
    \midrule
    $a/b$         & \uflc{a/b} \\
    $a^b$                 & \uflc{a**b}, \uflc{pow(a,b)} \\
    $\sqrt{f}$            & \uflc{sqrt(f)} \\
    $\exp f$              & \uflc{exp(f)} \\
    $\ln f$               & \uflc{ln(f)} \\
    $|f|$                 & \uflc{abs(f)} \\
    $\operator{sign} f$   & \uflc{sign(f)} \\
    \bottomrule
  \end{tabular}
  \hspace{0.5cm}
  \begin{tabular}{ll}
    \toprule
    Math. notation & UFL notation \\
    \midrule
    $\cos f$       & \uflc{cos(f)} \\
    $\sin f$       & \uflc{sin(f)} \\
    $\tan f$       & \uflc{tan(f)} \\
    $\arccos f$    & \uflc{acos(f)} \\
    $\arcsin f$    & \uflc{asin(f)} \\
    $\arctan f$    & \uflc{atan(f)} \\
    \\[0.14em]
    \bottomrule
  \end{tabular}
}
  \label{tab:elementarytrigfunctions}
\end{table}
\begin{table}
\tbl{
(Left) Table of special functions. (Right) Table of
    element-wise operators.
}{
  \begin{tabular}{ll}
    \toprule
    Math. notation & UFL notation \\
    \midrule
    $\operator{erf} f$        & \uflc{erf(f)} \\
    $J_{\nu}(f)$    & \uflc{bessel\_J(nu, f)} \\
    $Y_{\nu}(f)$    & \uflc{bessel\_Y(nu, f)} \\
    $I_{\nu}(f)$    & \uflc{bessel\_I(nu, f)} \\
    $K_{\nu}(f)$    & \uflc{bessel\_K(nu, f)} \\[0.19em]
    \bottomrule
  \end{tabular}
  \hspace{0.5cm}
  \begin{tabular}{ll}
    \toprule
    Math. notation & UFL notation \\
    \midrule
    $C \suchthat C_\alpha = A_\alpha B_\alpha$      & \uflc{elem\_mult(A, B)} \\
    $C \suchthat C_\alpha = A_\alpha / B_\alpha$    & \uflc{elem\_div(A, B)} \\
    $C \suchthat C_\alpha = A_\alpha^{B_\alpha}$      & \uflc{elem\_pow(A, B)} \\
    $C \suchthat C_\alpha = f(A_\alpha, \ldots)$  & \uflc{elem\_op(f, A, ...)} \\
    \\
    \bottomrule
  \end{tabular}
}
  \label{tab:specialfunctions}
\end{table}

\subsubsection{Differential operators and explicit variables}
\label{subsec:diffopandvariables}

UFL supports a range of differential operators. Most of these mirror
common ways of expressing spatial derivatives in partial differential
equations. A summary is presented in
Table~\ref{tab:differentialoperators} (left).

\begin{table}
\tbl{
(Left) Table of differential operators. (Right)
    Table of conditional operators.
}{
  \begin{tabular}{ll}
    \toprule
    Math. notation & UFL notation \\
    \midrule
    $A_{,i}$ or $\frac{\partial A}{\partial x_i}$  & \uflc{A.dx(i)} or
    \uflc{Dx(A, i)} \\
    $\frac{d A}{d n}$                  & \uflc{Dn(A)} \\
    $\Div A$                           & \uflc{div(A)} \\
    $\Grad A$                          & \uflc{grad(A)} \\
    $\nabla\cdot A$                    & \uflc{nabla\_div(A)} \\
    $\nabla A \equiv \nabla\otimes A$  & \uflc{nabla\_grad(A)} \\
    $\Curl A \equiv \nabla\times A$    & \uflc{curl(A)} \\
    $\operator{rot} A$                 & \uflc{rot(A)} \\
    $v = e$                            & \uflc{v = variable(e)} \\
    $\frac{d A}{d v}$                  & \uflc{diff(A, v)} \\
    $\dif f$  & \uflc{exterior\_derivative(f)} \\
    \bottomrule
  \end{tabular}
  \hspace{0.5cm}
  \begin{tabular}{ll}
    \toprule
    Math. notation & UFL notation \\
    \midrule
    $\begin{cases}T, \, \text{if} \, c\\ F, \, \text{otherwise}\end{cases}$       & \uflc{conditional(c, T, F)} \\
    $a = b$         & \uflc{eq(a, b)} \\
    $a \ne b$       & \uflc{ne(a, b)} \\
    $a \le b$       & \uflc{le(a, b)} or \uflc{a <= b} \\
    $a \ge b$       & \uflc{ge(a, b)} or \uflc{a >= b} \\
    $a < b$         & \uflc{lt(a, b)} or \uflc{a < b} \\
    $a > b$         & \uflc{gt(a, b)} or \uflc{a > b} \\
    $l \land r$     & \uflc{And(l, r)} \\
    $l \lor r$      & \uflc{Or(l, r)} \\
    $\lnot c$     & \uflc{Not(c)} \\
    \bottomrule
  \end{tabular}
}
  \label{tab:differentialoperators}
  \label{tab:conditionaloperators}
\end{table}

Basic partial derivatives, $\partial /\partial x_i$, can be
written as either \uflc{A.dx(i)} or \uflc{Dx(A,i)}, where
\uflc{i} is either a free-index or a fixed-index in the range
$[0,d)$. In the literature, there exist (at least) two conventions for
the gradient and the divergence operators depending on whether the
spatial derivative axis is appended or prepended; or informally,
whether gradients and divergences are taken column-wise or
row-wise. We denote the former convention by $\Grad(v)$ and the latter
by~$\nabla v$.  The two choices are reflected in UFL via two gradient
operators: \uflc{grad(A)} corresponding to $\Grad(v)$, and
\uflc{nabla\_grad(A)} corresponding to~$\nabla v$.  The two
divergence operators \uflc{div(A)} and \uflc{nabla\_div(A)} follow
the corresponding traditions. Also available are the curl operator and its
synonym rot, as well as a shorthand notation for the normal
derivative.

Expressions can also be differentiated with respect to user-defined
\emph{variables}. An expression \uflc{e} is annotated as a variable
via \uflc{v = variable(e)}. Letting \uflc{A} be some expression of
\uflc{v}, the derivative of \uflc{A} with respect to \uflc{v}
then reads as \uflc{diff(A, v)}. Note that~\uflc{diff(e, v) == 0},
because the expression \uflc{e} is not a function of \uflc{v},
just as $\frac{\dif f(x)}{\dif g(x)} = 0$ even if the equality
$f(x) = g(x)$ holds for two functions $f$ and $g$ since $f$ is defined
in terms of $x$ and not in terms of $g$.

\subsubsection{Conditional operators}
\label{sec:conditionaloperators}

The lack of control flow statements and mutable variables in UFL is
offset by the inclusion of conditional statements, which are
equivalent to the ternary operator known from, for example, the C
programming language.  To avoid issues with overloading the meaning of
some logical operators, named operators are available for all boolean
UFL expressions.  In particular, the equivalence operator \uflc{==}
is reserved in Python for comparing if objects are equal, and the
delayed evaluation behavior of Python operators \uflc{and},
\uflc{or} and \uflc{not} preclude their use as DSL components.  The
available conditional and logical operators are listed in
Table~\ref{tab:conditionaloperators} (right) and follow the standard
conventions.

\subsubsection{Discontinuous Galerkin operators}
\label{sec:dgoperators}

UFL facilitates compact implementation of discontinuous Galerkin
methods by providing a set of operators targeting discontinuous
fields. Discontinuous Galerkin discretizations typically rely on
evaluating jumps and averages of (discontinuous) piecewise functions,
defined relative to a tessellation, on both sides of cell facets. More
precisely, let $F$ denote an interior facet shared by the cells
$T^{+}$ and $T^{-}$, and denote the restriction of an expression $f$
to $T^{+}$ and $T^{-}$ by $f^{+}$ and $f^{-}$, respectively. In UFL,
the corresponding restrictions of an expression \uflc{f} are expressed
via \uflc{f('+')} and \uflc{f('-')}.

Two typical discontinuous Galerkin operators immediately derive from
these restrictions: the \emph{average} $\avg{f}=(f^{+} + f^{-})/2$,
and \emph{jump} operators~$\jump{f}=f^{+} - f^{-}$. For convenience,
these two operators are available in UFL via \uflc{avg(f)}
and~\uflc{jump(f)}. Moreover, it is common to use the outward unit
normal to the interior facet, $n$, when defining the jump operator
such that for a scalar-valued expression $\jump{f}_n = f^{+}n^{+} +
f^{-}n^{-}$, while for a vector- or tensor-valued expression
$\jump{f}_n=f^{+}\cdot n^{+} + f^{-}\cdot n^{-}$. These two
definitions are implemented in a single UFL operator \uflc{jump(f,
  n)} by letting UFL automatically determine the rank of the
expression and return the appropriate definition.  The available
operators are presented in Table~\ref{tab:dgoperators}~(left).

\begin{table}
\tbl{
(Left) Table of discontinuous Galerkin operators.
    (Right) Table of subdomain integrals.
}{\begin{tabular}{ll}
    \toprule
    Mathematical notation & UFL notation \\
    \midrule
    $f^{+}$, $f^{-}$    & \uflc{f('+')}, \uflc{f('-')} \\
    $\avg{f}$          & \uflc{avg(f)} \\
    $\jump{f}$         & \uflc{jump(f)} \\
    $\jump{f}_n$       & \uflc{jump(f, n)} \\
    \bottomrule
  \end{tabular}
  \hspace{0.5cm}
  \begin{tabular}{ll}
    \toprule
    Mathematical notation & UFL notation \\
    \midrule
    $\int_{\mathcal{T}_{k+1}} \,\,\, I \dif x$         & \uflc{I * dx(k)} \\
    $\int_{\mathcal{F}_{k+1}} I \dif s$ & \uflc{I * ds(k)} \\
    $\int_{\mathcal{F}^0_{k+1}} I \dif s$         & \uflc{I * dS(k)} \\[0.1em]
    \bottomrule
  \end{tabular}}
  \label{tab:dgoperators}
  \label{tab:integrals}
\end{table}

Since UFL is an embedded language, a user can easily implement custom
operators for discontinuous Galerkin methods from the basic
restriction building blocks. The reader is referred
to~\citet{oelgaard:2008} for more details on discontinuous Galerkin
methods in the context of a variational form language, developed for
FFC, and automated code generation.

\subsection{Integrals and variational forms}
\label{sec:forms}

In addition to expressions, UFL provides concepts and syntax for
defining integrals over facets and cells and, via integrals, for
defining variational forms. Variational forms can further be
manipulated and transformed via form operations. This sublanguage is
described in the sections below.

\subsubsection{Integrals and forms}

The integral functionality provided by UFL is centered around the
integrals featuring in variational forms as summarized by the
canonical expression~\eqref{eq:canonical}. An integral is generally
defined by an integration domain, an integrand and a measure. UFL
admits integrands defined in terms of terminal expressions and
operators as described in the previous section. The mathematical
concept of an integration domain together with a corresponding measure
is however embodied in a single abstraction in UFL, namely the
\uflc{Measure} class. This abstraction was inherited from the
original FFC form language.

A UFL measure (for simplicity just a measure from here onwards) is
defined in terms of a domain type, a domain identifier and,
optionally, additional domain meta-data. The allowed domain types
include \uflc{"cell"}, \uflc{"exterior\_facet"} and
\uflc{"interior\_facet"}. These domain types correspond to Lebesgue
integration over (a subset of) cells, (a subset of) exterior facets or
(a subset of) interior facets.  The domain identifier must be a
non-negative integer (the index~$k$ in~\eqref{eq:canonical}). For
convenience, three measures are predefined by UFL: \uflc{dx},
\uflc{ds} and \uflc{dS} corresponding to measures over cells,
exterior facets and interior facets, respectively, each with the
default domain identifier~$0$. However, new measures can be created
directly, or by calling a measure with an integer $k$ yielding a new
measure with domain identifier~$k$. Several other, less common, domain
types are allowed, such as \uflc{"surface"}, \uflc{"point"} and
\uflc{"macro\_cell"}, and new measures are easily added. We refer to
the UFL documentation for the complete description of these.

A UFL integral is then defined via an expression, acting as the
integrand, and a measure object. In particular, multiplying an
expression with a measure yields an integral. This is illustrated in
Table~\ref{tab:integrals} (right), with \uflc{k} denoting the domain
identifier. We remark that all integrand expressions featuring in an
integral over interior facets must be restricted for the integral to
be admissible. The integrand expression will depend on a number of
distinct argument functions: the number of such is labeled the arity
of the integral.

Finally, a UFL form is defined as the sum of one or more integrals. A
form may have integral terms of different arities. However, if all
terms have the same arity $\rho$ we term this $\rho$ the arity of the
form.  Forms are labeled according to their arity: forms of arity $0$
are called functionals, forms of arity $1$ are called linear forms, and
forms of arity $2$ are called bilinear forms. We emphasize that the
number of coefficient functions does not affect the arity of an
integral or a form. Table~\ref{tab:forms} shows simple examples of
functionals, linear forms and bilinear forms, while more complete
examples are given in Section~\ref{sec:examples}.

\begin{table}
\tbl{
Table of various form examples.
}{
  \begin{tabular}{lll}
    \toprule
    Mathematical notation & UFL notation & Form type \\[0.5em]

    \midrule
    $M(f;) = \int_{\mathcal{T}_1} \frac{1-f^2}{1+f^2} \dif x$ &
    \uflc{M = (1-f**2)/(1+f**2) * dx} &
    Functional \\[0.5em]

    $M(f, g;) = \int_{\mathcal{T}_1} \Grad f \cdot \Grad g \dif x$ &
    \uflc{M = dot(grad(f), grad(g)) * dx} &
    Functional \\[0.5em]

    $L(v) = \int_{\mathcal{T}_1} \sin(\pi x)v \dif x$ &
    \uflc{L = sin(pi*x[0])*v * dx} &
    Linear form \\[0.5em]

    $L(g; v) = \int_{\mathcal{F}_1} (\Grad g \cdot n) v \dif s$ &
    \uflc{L = Dn(g)*v * ds} &
    Linear form \\[0.5em]

    $a(u, v) = \int_{\mathcal{T}_1} u v \dif x$ &
    \uflc{a = u*v * dx} &
    Bilinear form \\[0.5em]

    $a(u, v) = \int_{\mathcal{T}_1} u v \dif x + \int_{\mathcal{F}_2} f u v \dif s $ &
    \uflc{a = u*v * dx(0) + f*u*v * ds(1)} &
    Bilinear form \\[0.5em]

    $a(u, v) = \int_{\mathcal{F}^0_1} \avg u \avg v \dif s$ &
    \uflc{a = avg(u)*avg(v) * dS} &
    Bilinear form \\[0.5em]

    $a(A; u, v) = \int_{\mathcal{T}_1} A_{ij} u_{,i} v_{,j} \dif x$ &
    \uflc{a = A[i,j]*u.dx(i)*v.dx(j) * dx} &
    Bilinear form \\[0.5em]

    \bottomrule
  \end{tabular}
}
\tabnote{With reference to coefficient
    functions \uflc{f}, \uflc{g}; argument functions \uflc{u},
    \uflc{v}; predefined integration measures \uflc{dx}, \uflc{ds},
    and \uflc{dS}; and subdomain notation as introduced in
    Section~\ref{subsubsec:geometric_domains_and_integrals}. Recall
    that the predefined UFL measures \uflc{dx}, \uflc{ds} and
    \uflc{dS} default to \uflc{dx(0)}, \uflc{ds(0)} and \uflc{dS(0)},
    respectively, and that the mathematical notation starts counting at
    $1$, while the code starts counting at $\uflc{0}$.}
  \label{tab:forms}
\end{table}

\subsubsection{Form operators}

UFL provides a select but powerful set of algorithms that act on forms
to produce new forms. An overview of such algorithms is presented and
exemplified in Table~\ref{tab:formoperators}.

\begin{table}
\tbl{
Overview of common form operators.
}{\scriptsize
  \begin{tabular}{cll}
    \toprule
    Mathematical notation & UFL notation & Description \\
    \midrule
    - & \uflc{L = lhs(F)}
    & Extract terms of arity $2$ from \uflc{F} \\
    - & \uflc{a = rhs(F)}
    & Extract terms of arity $1$ from \uflc{F}, multiplied with -1 \\
    - & \uflc{(a, L) = system(F)}
    & Extract both lhs and rhs terms from \uflc{F} \\
    $a \mapsto a^{\ast}$  & \uflc{a\_star = adjoint(a)}
    & Derive adjoint form of bilinear form \uflc{a} \\
    $F(f; \cdot) \mapsto F(g; \cdot)$ & \uflc{G = replace(F, \{f:g\})}
    & Replace coefficient \uflc{f} with \uflc{g} in \uflc{F} \\
    $F(; \cdot) \mapsto F(f; \cdot)$
    & \uflc{M = action(F, f)}
    & Replace argument function $1$ in \uflc{F} by \uflc{f} \\
    $F(f; \cdot) \mapsto D_f F(f; \cdot)[v]$
    & \uflc{dF = derivative(F, f, v)}
    & Differentiate \uflc{F} w.r.t \uflc{f} in direction \uflc{v} \\
    \bottomrule
  \end{tabular}
}
\tabnote{With reference to a
    given form \uflc{F} of possibly mixed arity, coefficient functions
    \uflc{f}, \uflc{g} and an argument function \uflc{v}. Note that
    all form operators return a new form as the result of the
    operation; the original form is unchanged.
}
  \label{tab:formoperators}
\end{table}

The first three operators in Table~\ref{tab:formoperators},
\uflc{lhs}, \uflc{rhs} and \uflc{system}, extract terms of certain
arities from a given form. More precisely, \uflc{lhs} returns the form
that is the sum of all integrals of arity~$2$ in the given form,
\uflc{rhs} returns the form that is the \emph{negative} sum of all
integrals of arity~$1$ and \uflc{system} extracts the tuple of both
the afore results; that is, \uflc{system(F)~=~(lhs(F),~rhs(F))}. These
operators are named by and typically used for extracting `left-hand'
and `right-hand' sides of variational equations expressed as $F(u, v)
= 0$. Next, the \uflc{adjoint} operator acts on bilinear forms to
return the adjoint form $a^{\ast}$: $a(;u, v) \mapsto a^{\ast}(;u, v)
= a(;v, u)$. The \uflc{replace} operator returns a version of a given
form in which given functions are replaced by other given
functions. The \uflc{action} operator can be viewed as special case of
replace: the argument function with the lowest count is replaced by a
given function.

The \Gateaux{} derivative~\eqref{eq:diffdef} is provided by the
operator \uflc{derivative}, and is possibly the most important form
operator. Table~\ref{tab:derivative} provides some usage examples of
this operator. With reference to Table~\ref{tab:derivative}, we
observe that forms can be differentiated with respect to coefficients
separately ($L_1$ and $L_2$) or with respect to simultaneous variation
of multiple coefficients~($L_3$). Note that in the latter case, the
result becomes a linear form with an argument function in the mixed
space. Differentiation with respect to a single component in a
vector-valued coefficient is also supported~($L_4$).

\begin{table}
\tbl{
Example derivative calls.
}{
  \begin{tabular}{ll}
    \toprule
    Mathematical operation & UFL notation \\
    \midrule
  $L_1(u, p; v) = \frac{d}{d\tau} \left[ M(u + \tau v, p)\right]_{\tau=0}$
  &
  \uflc{L1 = derivative(M, u, v)}
  \\
  $L_2(u, p; q) = \frac{d}{d\tau} \left[ M(u, p + \tau q) \right]_{\tau=0}$
  &
  \uflc{L2 = derivative(M, p, q)}
  \\
  $L_3(u, p; w) =  \frac{d}{d\tau} \left[ M(u + \tau v, p + \tau q) \right]_{\tau=0}$
  &
  \uflc{L3 = derivative(M, (u, p), w)}
  \\
  $L_4(u, p; s) =  \frac{d}{d\tau} \left[ M(u + \tau s e_y, p) \right]_{\tau=0}$
  &
  \uflc{L4 = derivative(M, u[1], s)}
  \\
    \bottomrule
  \end{tabular}
}
\tabnote{
With reference to a functional $M: W = V \times Q
    \rightarrow \R$ for a vector-valued function space $V$ with a
    scalar subspace $V_1$, and a scalar-valued function space $Q$;
    coefficient functions $u \in V$, $p \in Q$; and
    argument functions $v \in V$, $q \in Q$, $s \in V_1$ and $w \in
    W$.}
  \label{tab:derivative}
\end{table}

The high-level operations on forms provided by UFL can enable the
expression of algorithms at a higher abstraction level than what is
possible or practical with a traditional implementation. Some concrete
examples using UFL and operations on forms can be found
in~\citep{RognesLogg201x,FarrellEtAl201x}.

\subsection{The \uflc{.ufl} file format}

UFL may be integrated into a problem solving environment in Python or
written in \uflc{.ufl} files and compiled offline for use in a problem
solving environment in a compiled language such as C++.  The
\uflc{.ufl} file format is simple: the file is interpreted by Python
with the full \uflc{ufl} namespace imported, and forms and elements
are extracted by inspecting the resulting namespace.  In a typical
\uflc{.ufl} file, only minor parts of the Python language are used,
although the full language is available. It may be convenient to use
the Python \uflc{def} statement to define reusable functions within a
\uflc{.ufl} file. By default, forms with the names \uflc{a}, \uflc{L},
\uflc{M}, \uflc{F} or \uflc{J} are exported (i.e. compiled by the form
compiler).  The convention is that \uflc{a} and \uflc{L} define
bilinear and linear forms for a linear equation, \uflc{M} is a
functional, and \uflc{F} and \uflc{J} define a nonlinear residual form
and its Jacobian. Forms with any names may be exported by defining a
list \uflc{forms = [form0, form1]}. By default, the elements that are
referenced by forms are compiled, but elements may also be exported
without being used by defining a list \uflc{elements = [element0,
    element1]}.  Note that meshes and values of coefficients are
handled by accompanying problem solving environment (such as DOLFIN)
and \uflc{Coefficient} and \uflc{Constant} instances in \uflc{.ufl}
files are purely symbolic.

\section{Examples}
\label{sec:examples}

We now present a collection of complete examples illustrating the
specification, in UFL, of the finite element discretizations of a
number of partial differential equations. We have chosen problems that
compactly illustrate particular features of UFL. We stress, however,
that UFL is not limited to simple equations. On the contrary, the
benefits of using UFL are the greatest for complicated, non-standard
equations that cannot be easily or quickly solved using conventional
libraries. Computational results produced using UFL have been
published in many works, covering a vast range of fields and problem
complexity, by third-parties and by the developers of UFL, including
~\citep{abert2012numerical,arnold2012consistency,brandenburg2012advanced,brunner2012deterministic,FunkeFarrell2013,hake2012modelling,labeur:2012,fenics:book,maraldi:2011,mortensen:2011,rosseel:2012,wells:2011} to mention but a few.

\subsection{Poisson equation}

As a first example, we consider the Poisson equation and its
discretization using the standard $H^1$-conforming formulation, a
$L^2$-conforming formulation and a mixed $H(\mathrm{div})/L^2$-conforming
formulation. The Poisson equation with boundary conditions is given by
\begin{equation}
  \label{eq:poisson}
  - \Div (\kappa \Grad u) = f \quad  \text{in } \Omega,
  \qquad
  u = u_0 \quad \text{on } \GammaD,
  \qquad
  -\kappa \partial_n u = g \quad \text{on } \GammaN,
\end{equation}
where $\GammaD \cup \GammaN = \partial\Omega$ is a partitioning of the
boundary of $\Omega$ into Dirichlet and Neumann boundaries,
respectively.

\subsubsection{$H^1$-conforming discretization}

\begin{figure}
  \centering
  \includegraphics[width=0.5\textwidth]{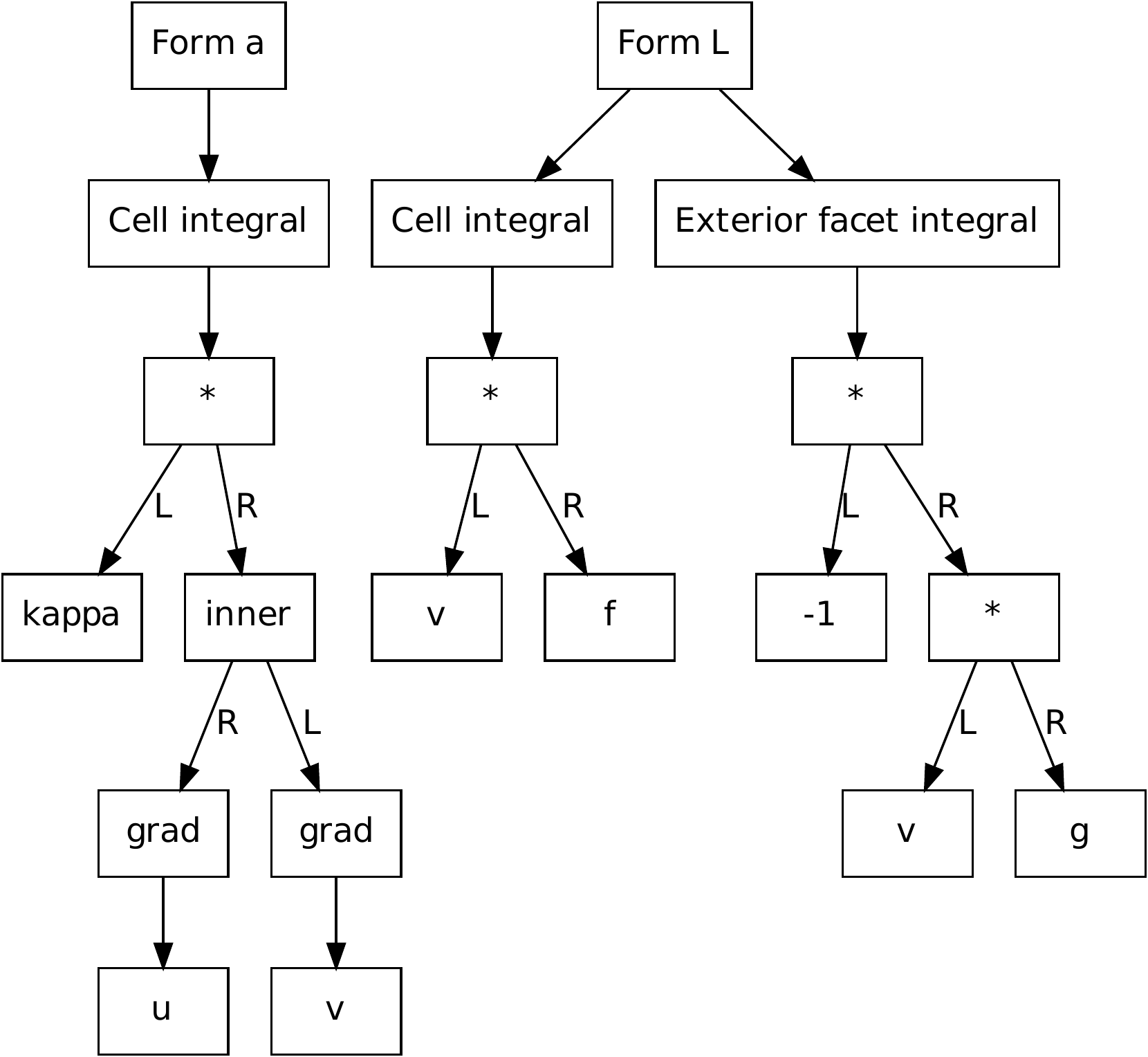}
  \caption{Expression trees for the $H^1$ discretization of the
    Poisson equation~\eqref{eq:poisson,H1}.}
  \label{fig:GraphPoissonH1}
\end{figure}

The standard $H^1$-conforming finite element discretization
of~\eqref{eq:poisson} reads: find $u \in V_h$ such that
\begin{equation}
  \label{eq:poisson,H1}
  a(u, v) \equiv \int_{\Omega} \kappa \Grad u \cdot \Grad v \dx
  =
  \int_{\Omega} f v \dx - \int_{\GammaN} gv \ds \equiv L(v)
\end{equation}
for all $v \in \hat{V}_h$, where $V_h$ is a continuous piecewise
polynomial trial space incorporating the Dirichlet boundary conditions
on $\GammaD$ and $\hat{V}_h$ is a continuous piecewise polynomial test
space with zero trace on $\GammaD$. We note that as a result of the
zero trace of the test function $v$ on $\GammaD$, the boundary
integral in~\eqref{eq:poisson,H1} may be expressed as an integral over
the entire boundary $\partial\Omega$. A complete specification of the
variational problem~\eqref{eq:poisson,H1} in UFL is included
below. The resulting expression trees for the bilinear and linear
forms are presented in~Figure~\ref{fig:GraphPoissonH1}.

\inputufl{examples/poisson/PoissonH1.ufl}

\subsubsection{$L^2$-conforming discretization}

For the $L^2$ discretization of the Poisson equation, the standard
discontinuous Galerkin/interior penalty
formulation~\citep{Arnold1982,oelgaard:2008} reads: find $u \in V_h$
such that
\begin{equation}
  \label{eq:poisson,L2}
  \begin{split}
     \int_\Omega \kappa \Grad u &\cdot \Grad v \dx
     - \int_{\GammaD} \kappa \partial_n u v \ds
     - \int_{\GammaD} \kappa \partial_n v u \ds
     + \int_{\GammaD} \frac{\gamma\kappa}{h} uv \ds \\
     + \sum_{F\in\mathcal{F}^0} \Bigg(
    &- \int_F \avg{\kappa \Grad u} \cdot \jump{v}_n \ds
     - \int_F \avg{\kappa \Grad v} \cdot \jump{u}_n \ds
     + \int_F \frac{\gamma \avg{\kappa}}{\avg{h}}\jump{u}\jump{v} \ds \Bigg) \\
    &\quad
     = \int_{\Omega} fv \dx
     - \int_{\GammaN} gv \ds
     - \int_{\GammaD} \partial_n u_0 v \ds
     - \int_{\GammaD} \partial_n v u_0 \ds
     + \int_{\GammaD} \frac{\gamma\kappa}{h} u_0 v \ds
  \end{split}
\end{equation}
for all $v \in V_h$,
where $\jump{v}$, $\jump{v}_n$ and $\avg{v}$ are the standard jump,
normal jump and average operators (see Section~\ref{sec:dgoperators}).

\begin{sidewaysfigure}
  \centering
  \includegraphics[width=\textwidth]{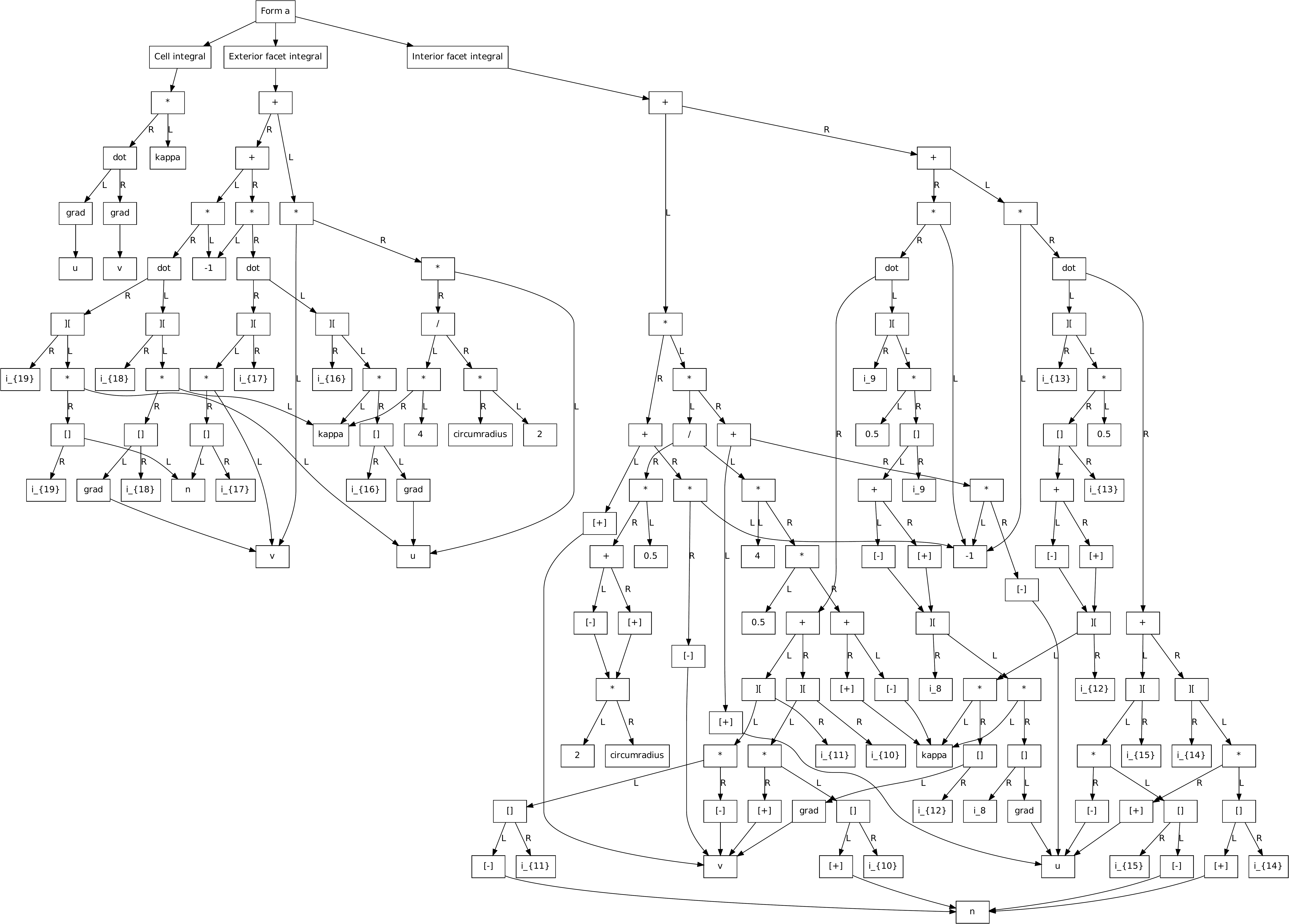}
  \caption{Expression tree for the $L^2$ discretization of the Poisson
    equation~\eqref{eq:poisson,L2}.  This expression tree serves
    as an illustration of the complexity of the expression tree already
    for a moderately simple formulation like the $L^2$ discretization
    of the Poisson equation. A detail of this expression tree is plotted
    in Figure~\ref{fig:GraphPoissonL2zoom}.}
  \label{fig:GraphPoissonL2}
\end{sidewaysfigure}

\begin{figure}
  \centering
  \includegraphics[width=0.3\textwidth]{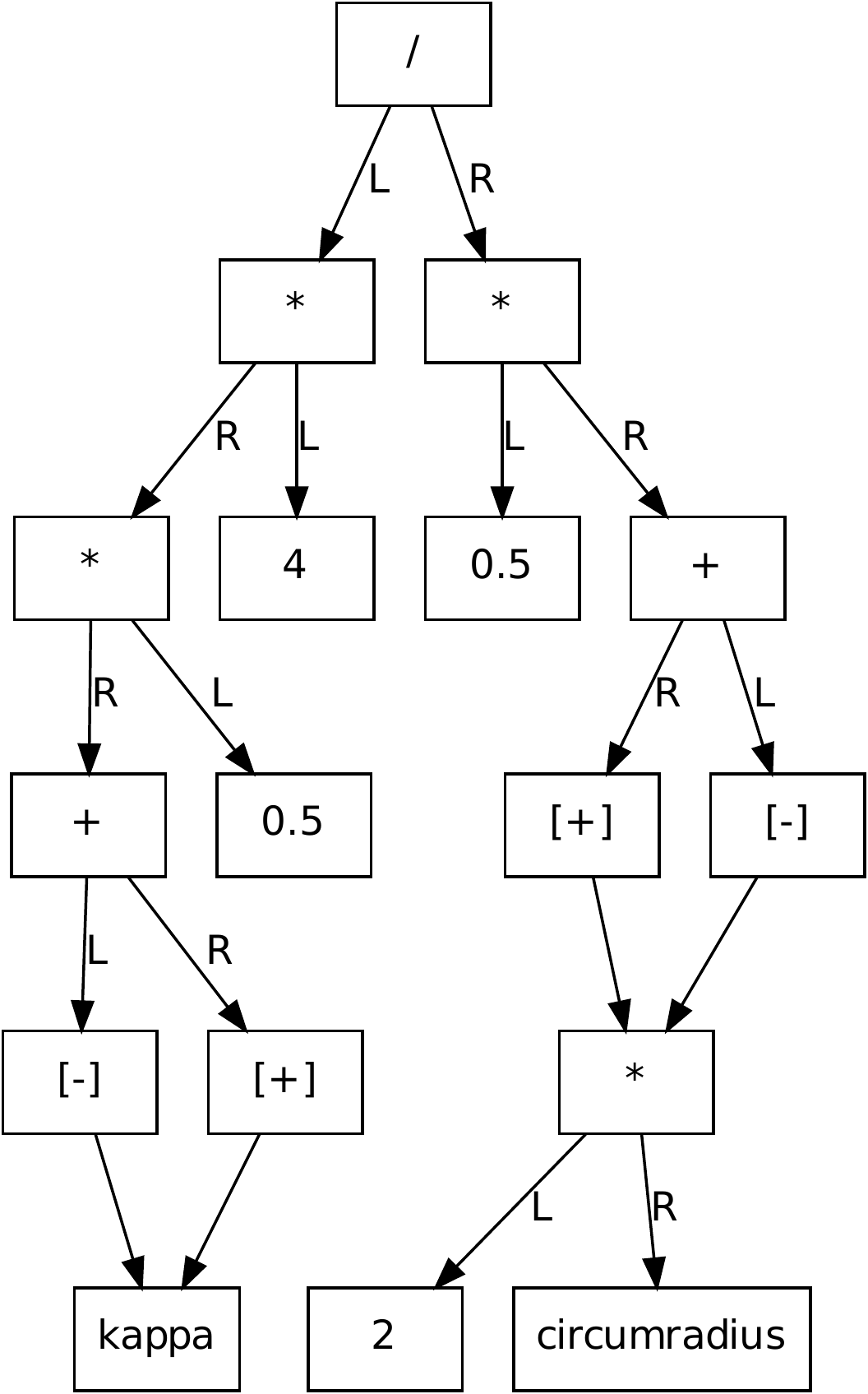}
  \caption{Detail of the expression tree of
    Figure~\ref{fig:GraphPoissonL2} for the $L^2$ discretization of
    the Poisson equation~\eqref{eq:poisson,L2}. The expression tree
    represents the expression \uflc{gamma*avg(kappa)/avg(h)}.}
  \label{fig:GraphPoissonL2zoom}
\end{figure}

The corresponding implementation in UFL is shown below. The relevant
operators for the specification of discontinuous Galerkin methods are
provided by the operators \uflc{jump()} and \uflc{avg()}. We also
show in Figures~\ref{fig:GraphPoissonL2}
and~\ref{fig:GraphPoissonL2zoom} the expression tree for the bilinear
form of the $L^2$-discretization, which is notably more complex than
for the $H^1$-discretization.

\inputufl{examples/poisson/PoissonL2.ufl}

\subsubsection{$H(\mathrm{div})/L^2$-conforming discretization}

Finally, we consider the discretization of the Poisson equation with a
mixed formulation, where the second-order PDE in~\eqref{eq:poisson} is
replaced by a system of first-order equations:
\begin{equation}
  \label{eq:mxpoisson}
              \Div \sigma = f     \quad \text{in } \Omega,  \qquad
  \sigma + \kappa \Grad u = 0     \quad \text{in } \Omega,  \qquad
                        u = u_0   \quad \text{on } \GammaD, \qquad
           \sigma \cdot n = g     \quad \text{on } \GammaN.
\end{equation}
A direct discretization of the system~\eqref{eq:mxpoisson} with, say,
continuous piecewise linear elements is unstable. Instead a suitable
pair of finite element spaces must be used for $\sigma$ and $u$. An
example of such a pair of elements is the $H(\mathrm{div})$-conforming
BDM (Brezzi--Douglas--Marini) element~\citep{BrezziDouglasMarini1985}
of the first degree for~$\sigma$ and discontinuous piecewise constants
for~$u$. Multiplying the two differential equations
in~\eqref{eq:mxpoisson} with test functions~$\tau$ and $v$,
respectively, integrating by parts and summing the variational
equations, we obtain the following variational problem: find $(\sigma,
u) \in V_h = V_h^{\sigma} \times V_h^u$ such that
\begin{equation}
  \int_{\Omega} (\Div \sigma) v \dx +
  \int_{\Omega} \sigma \cdot \tau \dx -
  \int_{\Omega} u \Div (\kappa\tau) \dx
  =
  \int_{\Omega} fv \dx -
  \int_{\Gamma_D} \kappa u_0 \tau \cdot n \ds
\end{equation}
for all $(\tau, v) \in \hat{V}_h$. Note that the Dirichlet condition
$u = u_0$ on $\GammaD$ is a \emph{natural} boundary condition in the
mixed formulation; that is, it is naturally imposed as a weak boundary
condition as part of the variational problem. On the other hand, the
Neumann boundary condition $\sigma \cdot n = g$ must be imposed as an
essential boundary condition as part of the solution space~$V_h$. The
corresponding specification in UFL is given below.

\inputufl{examples/poisson/PoissonHdivL2.ufl}

\subsection{Stokes equations}

As a second example of a mixed problem, we consider the Stokes
equations given by
\begin{equation}
  -\Delta u + \Grad p = f \quad \text{in } \Omega, \qquad
               \Div u = 0 \quad \text{in } \Omega,
\end{equation}
for a velocity u and a pressure p,
together with a suitable set of boundary conditions. The first
equation is multiplied with a test function $v$, the second equation
is multiplied with a test function $q$, and after integration by
parts, the resulting mixed variational problems is: find $(u, p) \in
V_h$ such that
\begin{equation}
  \int_{\Omega} \Grad u \cdot \Grad v \dx -
  \int_{\Omega} (\Div v) p \dx +
  \int_{\Omega} (\Div u) q \dx
  =
  \int_{\Omega} fv \dx.
\end{equation}
The corresponding specification in UFL using an
$\mathrm{inf}$--$\mathrm{sup}$ stable $[P_2]^d$--$P_1$ discretization
(Taylor--Hood element~\citep{TaylorHood1973}) is shown below.

\inputufl{examples/stokes/Stokes.ufl}

\subsection{Neo-Hookean hyperelastic model}

We consider next a hyperelastic problem posed in terms of the
minimization of potential energy. For a body $\Omega \subset
\mathbb{R}^{d}$ in a reference configuration, where $1 \le d \le 3$,
the total potential energy $\Pi$ reads
\begin{equation}
  \Pi(v) = \int_{\Omega} \psi(v) \dx - \int_{\Omega} B \cdot v \dx
  - \int_{\Gamma} T \cdot  v \ds,
\label{eqn:potential_energy}
\end{equation}
where $\psi$ is the stored energy density, $v$ is the displacement
field, $B$ is the nominal body force and $T$ is the nominal traction
on the domain boundary $\Gamma = \partial \Omega$. For the
compressible neo-Hookean model, the strain energy density reads
\begin{equation}
  \psi(v) =
  \frac{\mu}{2}(I_c - 3) - \mu\ln J + \frac{\lambda}{2}(\ln J)^2,
\end{equation}
where $J$ is the determinant of the deformation gradient $F = I +
\Grad u$ and $I_c$ is the trace of the right Cauchy--Green tensor $C =
F^{T}F$. Solutions $u$ to the hyperelastic problem
minimize~\eqref{eqn:potential_energy}:
\begin{equation} \label{eq:minproblem}
  u = \underset{v \in V}{\operatorname{argmin}} \, \Pi(v).
\end{equation}
The classical variational approach to solving~\eqref{eq:minproblem}
involves taking the first variation/\Gateaux{} derivative of
\eqref{eqn:potential_energy} with respect to $v$ and setting this equal
to zero for all~$v$, which in turn can be solved using Newton's method
by taking a second \Gateaux{} derivative to yield the Jacobian.  The
corresponding specification in UFL for such an approach is shown below.

\inputufl{examples/hyperelasticity/HyperElasticity.ufl}

A complete solver using precisely the above formulation is distributed
as a demo program as part of DOLFIN~\citep{logg:2010,dolfin:www}.

\subsection{Constrained optimization}

As a final example, we consider a PDE constrained optimization
problem. We wish to determine the control variable $p$ that minimizes
the cost functional
\begin{equation}
  J(u, p) =
  \int_\Omega \frac{1}{2}(u - \bar{u})^2 \dx +
  \int_\Omega\frac{1}{2}\alpha p^2 \dx,
\end{equation}
where $\bar{u}$ is a given target function, $\alpha$ is a
regularization parameter and $u$ is the state, constrained by the
variational problem
\begin{equation}
  \label{eq:optimization,varproblem}
  a(u, v) \equiv \int_\Omega u v + \Grad u \cdot \Grad v \dx
  = \int_\Omega p v \dx \equiv b(p; v),
\end{equation}
which should hold for all test functions~$v$ in some suitable test space.

A way to solve the optimization problem is to define the Lagrangian
functional $\mathcal{L}$ as the sum of the cost functional and the
weak constraint, in which $v$ plays the role of a Lagrange multiplier:
\begin{equation}
  \mathcal{L}(u, p, v) = J(u, p) + a(u, v) - b(p; v).
\end{equation}
The solution of the optimization problem can be found by seeking the
stationary points of the Lagrangian:
\begin{alignat}{2}
  D_u &\mathcal{L}(u, p, v)[\delta u] &= 0 \qquad & \forall \ \delta u, \\
  D_p &\mathcal{L}(u, p, v)[\delta p] &= 0 \qquad & \forall \ \delta p, \\
  D_v &\mathcal{L}(u, p, v)[\delta v] &= 0 \qquad & \forall \ \delta v,
\end{alignat}
which in this case is a linear system of equations. The corresponding
implementation in UFL is shown below.

\inputufl{examples/optimization/Optimization.ufl}

\section{Representation of expressions}
\label{sec:representation}

UFL is a collection of value types and operators. These types and
operators have been presented in the preceding sections from a user
perspective, with a focus on their mathematical definitions.  To
discuss the representation and algorithms in the symbolic framework
underlying the language implementation, we will now pursue a more
abstract approach.

\subsection{Abstract categorization of expression types}

As a domain specific language embedded in Python, UFL does not have a
formal grammar of its own. However, it is still useful to compactly
categorize the various types of expressions before proceeding. An
expression $e$ can be either a terminal expression $t$, or an operator
$o$ applied to one or more other expressions. A terminal expression
$t$ can be categorized in one of the groups:
\begin{benumerate}
  \bolditem[$i$] Multi-index.
  \bolditem[$l$] Literal constant.
  \bolditem[$g$] Geometric quantity.
  \bolditem[$f$] Function. Further classified as:
  \begin{benumerate}
    \bolditem[$c$] Coefficient function.
    \bolditem[$a$] Argument function.
  \end{benumerate}
\end{benumerate}
An operator type $o$ can be categorized in one of the following
groups:
\begin{benumerate}
  \bolditem[$I$] Indexing related operators, which manipulate the
  indices and shapes of expressions. These operations do not introduce
  any computation.

  \bolditem[$A$] Basic arithmetic operators, including the operators \uflc{+},
  \uflc{-}, \uflc{*} and~\uflc{/}.

  \bolditem[$F$] Nonlinear scalar real functions, such as $\ln(x)$,
  $\exp(x)$, $\sin(x)$ and~$\arccos(x)$.

  \bolditem[$T$] Tensor algebraic operators. These are convenience
  operators acting on nonscalar expressions, such as the dot, inner,
  and outer products, and the transpose, determinant or deviatoric part.

  \bolditem[$D$] Differential operators, including both spatial
  derivatives, derivatives with respect to expressions annotated as
  variables and directional derivatives with respect to coefficient
  functions.

  \bolditem[$R$] Restrictions of functions to cells and related operators
  such as the jump and average between cells.

  \bolditem[$B$] Boolean operators, including $=$, $\ne$, $<$, $>$,
  $\le$, $\ge$, $\land$, $\lor$, $\lnot$. These can only be used within
  a conditional operator.

  \bolditem[$C$] The conditional operator: the ternary operation ``$f$ if
  condition else $g$''. This is the only explicit branching instruction,
  not counting restrictions.

\end{benumerate}
With these definitions, the following diagram gives a compact
semi-formal overview of the types and operators that an expression $e$
can be recursively built from:
\begin{align*}
  t &= [i | l | g | c | a], \\
  o &= [I | A | F | T | D | R | B | C], \\
  e &= [t | o(e_1, \ldots, e_n)].
\end{align*}
As before, $t$ represents any terminal expression, while $o$
represents any operator.  If an expression $e$ is an application of
an operator, we say that it \emph{depends} on the expressions $e_1, \ldots, e_n$.
Dependencies between expressions are, by construction, one-way relations
and can therefore be viewed as the edges of a directed acyclic
graph (DAG), where each vertex is itself an expression~$e_i$. Each
graph vertex has an associated type, which is either a terminal type or
operator type. Vertices representing a terminal expression may have
additional descriptive data associated with them, while vertices
representing non-terminal subexpressions are fully identified by the
type and sequence of child vertices. We note here that the ordering of
child vertices (dependencies) is important (for nonsymmetric operators).

\subsection{Representation of directed acyclic graphs}

From an implementation viewpoint, a DAG can be represented by
link-based or list-based data structures.  In a link-based data
structure, each vertex of the graph is represented by a single typed
object. Objects of operator types store references to the objects
representing their operands. These references are the edges of the
DAG.  Objects of terminal types may store additional data. This is the
natural DAG representation for a symbolic library while an expression
is being built, and is the way UFL expressions are represented most of
the time. When a UFL operator is evaluated in Python, the result is a
new object\footnote{Sometimes more than one new object, but the number
  of auxiliary objects is always bounded by a small constant.}
representing the operator. Thus, both the time and storage cost of
building an expression is $O(n)$ in the number $n$ of operators
applied.  Since each operator invocation leads to a new graph
vertex without any global knowledge available, some duplicate
expressions may occur in this DAG representation.

A list-based data structure for an expression DAG can be constructed
from the link-based data structure, if needed. In this data structure,
references to each unique subexpression are placed in a topologically
sorted vertex list and all edges (dependencies) are stored in another
list as a tuple of vertex list indices. While the vertex list is
constructed, duplicate expressions with the exact same symbolic
representation are automatically mapped to the same vertex.  This
mapping can be achieved through $O(1)$ insertion into a hash map,
which retains the overall $O(n)$ performance of building this data
structure. The main advantages of the list-based graph representation
are efficient ordered iteration over all vertices and easy access to
dependency information. This is beneficial for some algorithms.

\subsection{Expression type class hierarchies}

Each DAG vertex, or expression object, is an instance of a subclass of
\uflc{Expr}. The type of terminal or operator is determined by the
subclass, from a class hierarchy which is divided into
\uflc{Terminal} and \uflc{Operator} subtypes. Concrete
\uflc{Operator} subclasses store references to the objects
representing their operands. Since a non-terminal expression is
uniquely determined by its type and operands, any other data stored by
these classes is purely for performance or convenience reasons.
Concrete \uflc{Terminal} subclasses, however, can take any necessary
auxiliary data in their constructor. This data must still be
immutable, such that any expression object is hashable. Most operator
overloading is applied directly to the \uflc{Expr} class. For
example, adding any two expressions will result in a new instance of a
\uflc{Sum}. All expression classes must overload a set of basic
methods. One such method is \uflc{operands()}, which returns a tuple
of expression objects for the operands of an \uflc{Operator}, or the
empty tuple for a \uflc{Terminal} object. We refer to the UFL source
code~\citep{ufl:www} for further internal implementation details.

\subsection{Representation of indexed expressions}

In some symbolic libraries, expressions with shapes or indices are
built on top of an otherwise mainly scalar framework. In contrast, UFL
considers shapes and indices as an integral part of any expression. To
support this, expressions provide three methods, namely,
\uflc{shape()} which returns a tuple of positive integers
representing the value shape of the expression,
\uflc{free\_indices()} which returns an unordered tuple of
\uflc{Index} objects for each free-index in the expression and
\uflc{index\_dimensions()}, which returns a mapping from
\uflc{Index} objects to the dimensions they implicitly range
over. Together, these methods provide a rich and flexible description
of the value shape and free-index set. These properties generalize to
arbitrarily shaped expressions with arbitrary sets of free-indices, and
are accessible for any expression regardless of type.

In traditional programming languages, the indexing operator \uflc{[]}
usually extracts a value from a container. In contrast, in a symbolic
environment, the result may be an object representing the operation of
extracting a component. When indexing with free-indices, the former
type of extraction is not possible since the index represents a range
of values. Instead, when indexing a tensor-valued expression in UFL,
the result is represented as an object of the \uflc{Indexed} class,
with the original expression and a \uflc{MultiIndex} object as
operands. A \uflc{MultiIndex} object\footnote{The type system in UFL
  is fairly simple, and \uflc{MultiIndex} is one of the few exceptions
  in the \uflc{Expr} type hierarchy that does not represent a
  tensor-valued expression.}  represents a sequence of \uflc{Index}
and \uflc{FixedIndex} objects. The opposite operation from indexing,
mapping an expression with free-indices to be seen as a tensor-valued
expression, cf.~Section~\ref{subsubsec:index_notation} and in
particular Table~\ref{tab:indexingoperators}, is represented with the
\uflc{ComponentTensor} class, which similarly has the original
expression as well as a \uflc{MultiIndex} as operands. Expressions
which contain layers of mapping back-and-forth between index and
tensor notation may appear complex, but a form compiler can reduce
this complexity during the translation to low-level code.

The implicit summation notation in UFL is applied early in the
application of the \uflc{*} operator and the derivative operator
\uflc{.dx(i)}. Subsequently the summation is represented as explicit
sums over free-indices. This way, the implicit summation rules need
no special consideration in most algorithms. For example, the sum $u_i
v_i \equiv \sum_i u_i v_i$ is expressed in UFL code as
\uflc{u[i]*v[i]}, but is represented in the DAG as
\uflc{IndexSum(Product(u[i], v[i]), (i,))}\footnote{Here, the
  representations of \uflc{u[i], v[i], (i,)} are simplified for
  compact presentation.}.  A \uflc{Product} object can be
constructed directly in algorithms where implicit summation is not the
wanted behavior. When interpreting an \uflc{IndexSum} object in
symbolic algorithms, the range of the sum is defined by the underlying
summand expression; this is possible as any expression in UFL knows
its own shape, free-indices and dimensions of its free-indices.

\subsection{Simplification of expressions}

Automatic simplification of expressions is a central design issue for any
symbolic framework. At one end of the design spectrum are conservative
frameworks that preserve expressions exactly as written by the user,
analogous to an abstract syntax tree of the program input to a
compiler. At the other end of the spectrum are frameworks where all
expressions are transformed to a canonical form at construction time.
The design of UFL is guided by the intention to be the front end to
multiple form compilers, and is therefore fairly conservative. If a form
compiler does no rewriting, the generated code should transparently
resemble the UFL expressions as authored by the end-user.  However,
UFL is also a collection of symbolic algorithms, and the performance
and memory footprint of such algorithms can be significantly improved by
certain automatic simplifications of expressions at construction time,
if such simplifications reduce symbolic expression growth.

Because form compilers may employ different expression rewriting
strategies, we wish to avoid doing any simplifications that might remove
information a form compiler could make use of.  A canonical sorting of
two sum operands eases the detection of common subexpressions, such as
in the expression rewriting $(b + a)(a + b) \rightarrow (a + b)(a +
b)$.  However, a canonical sorting of more than two sum operands may
hide common subexpressions, such as in the rewriting $((a + c) + b)(a
+ c) \rightarrow (a + b + c)(a + c)$.  Therefore, sums and products
are represented in UFL as binary operators (as opposed to storing
lists of terms or factors), with their two operands sorted
canonically.  Even more importantly, we avoid any symbolic rewriting
that can lead to numerically unstable floating point expressions in
generated code.  An example of such unsafe operations is the expansion
of a factored polynomial $(a - b)(a - b) \rightarrow a^2 - 2ab + b^2$,
which becomes numerically unstable in inexact floating point
arithmetic.

The performance of the symbolic algorithms and the form compilation
processes poses a final limitation on the type of automatic
simplifications we may apply. It is crucial that the overall form
compilation process can be designed to have an asymptotic cost of $O(n)$
in time and memory usage\footnote{Form compilers are free to
  apply more expensive strategies, but UFL must not render efficient
  algorithms impossible.}, with $n$ a measure of the length of the
integrand expression. Because simplification of expressions at
construction time is performed once for each expression object, the cost
of applying the simplification must be a local operation independent
of the size of the operand expressions. Thus, automatic simplifications
which would involve traversing the entire subexpression DAG for analysis
or rewriting are never attempted.

The following small set of safe and local simplifications is applied
consistently when constructing expressions:
\begin{description}
\item[Multiply by one]
  $1 x \rightarrow x$.
  A simplification which keeps one operand intact and throws away the
  other one is safe.

\item[Add zero]
  $0 + x \rightarrow x$.

\item[Multiply by zero]
  $0 x \rightarrow 0$.
  To avoid losing the tensor properties of $x$, we annotate the
  representation of $0$ with the same properties. Therefore we have a
  special zero representation with shape and free-indices.

\item[Constant folding]
  $f(l_1, l_2) \rightarrow l_3$.
  Here $l_1$, $l_2$ and $l_3$ are literal constants and $f$ is a function
  or operator that can be computed numerically by UFL. This happens
  recursively and so constant scalar expressions are effectively folded
  to a single literal constant.

\item[Canceling indexing]
  $A_\alpha E^\alpha \rightarrow A$.
  The mappings between tensor-valued and indexed expression
  representations cancel when the same multi-index $\alpha$ is used in
  both operations. If the inner and outer multi-indices are not equal,
  this canceling operation would require rewriting the representation
  of A, which is not a local operation, and therefore not invoked.
\end{description}
Note that some of these operations will occur frequently during automatic
differentiation. For example, consider the derivative $d(fg) / df$,
where $f$ and~ $g$ are independent functions:
\begin{align}
  \frac{d(fg)}{df}
      = \frac{df}{df} g + f \frac{dg}{df}
      = 1 g + f 0 \rightarrow g + 0 \rightarrow g.
\end{align}
Similarly, during some symbolic algorithms, tensor-valued
subexpressions are indexed to simplify computation and later mapped back to tensor-valued
expressions. This process leads to superfluous indexing patterns which
causes the DAG to grow needlessly. Applying the indexing cancellations
helps in avoiding this DAG growth.

\section{Algorithms}
\label{sec:algorithms}

The UFL implementation contains a collection of basic algorithms for
analysis and transformation of expressions. These algorithms include
optimized Python generators for easy iteration over expression nodes
in pre- or post-ordering and iteration over terminal expressions or,
more generally, expressions of a particular type. In the following,
some of the core algorithms and building blocks for algorithms are
explained. Particular emphasis is placed on the differentiation
algorithm. To avoid verbose technical details in the discussions of
symbolic algorithms, we first define some mathematical notation to
support abstract algorithm descriptions. For implementation details,
we refer to the UFL source code~\citep{ufl:www}.

\subsection{Evaluation algorithm}

Assume that an expression $e$ is represented by a DAG with $m$
terminal and $n$ non-terminal subexpressions, recalling that each
subexpression is a DAG vertex. Denote the terminal subexpressions by $e_i$, for
$i = 1, 2, \ldots, m$ and the non-terminal subexpressions by $e_i$, for
$i = m + 1, m + 2, \ldots, m + n$. For each $e_i$, let $I^i \equiv \langle I^i_j
\rangle_{j=1}^{p_i}$ be a sequence of $p_i$ integer labels referring
to the operand expressions of $e_i$. Note that for each terminal
expression, this sequence is empty. Moreover, we require the labels to
fulfill a topological ordering such that
\begin{align}
  I^i_j < i, \qquad j = 1, \ldots, p_i, \qquad i = 1,
  \ldots, m+n.
\end{align}
We thus have $k < i$ whenever $e_i$ depends on $e_k$ (directly or indirectly).
Equivalently $e_i$ is independent of any subexpression $e_k$ for~$k > i$.

We can now formulate two versions of an algorithm for evaluating $e
\equiv e_{m+n}$.  Note that these algorithms are merely abstract tools
for describing the mathematical structure of algorithms that follow.
For each specific algorithm, we require an evaluation operator $E$
where $v_i = E(e_i)$ is called the value of $e_i$. This value can be a
floating point value, a new symbolic expression or a generated source
code string, depending on the purpose of the algorithm. The
implementation of the evaluation operator $E$ will in general depend
on the UFL expression type of its argument $e$, which we denote~$\type(e)$.

First, partition the set $T$ of all UFL expression types into disjoint
sets of terminal types $T_T$ and operator types $T_O$, and let $E_T$
and $E_O$ be non-recursive evaluation operators for terminal and
non-terminal expressions, respectively. We can then design a simple
recursive evaluation algorithm $E_R$ of the form
\begin{align}
  \label{eq:recursiveevaluateoperator}
  E_R(e_i) &= \begin{cases}
    E_T(e_i),
    \qquad & \text{if} \, \type(e_i) \in T_T, \\
    E_O\left(e_i, \langle E_R(e_{I^i_j}) \rangle_{j=1}^{p_i}\right),
    \qquad & \text{if} \, \type(e_i) \in T_O.
  \end{cases}
\end{align}
The algorithmic structure of~\eqref{eq:recursiveevaluateoperator}
assumes that any subexpression can be evaluated given the values of
its operands, which is not true for operators which provide a
context for the evaluation of their operands. For example, in a
derivative evaluation algorithm, each type of derivative operator
provides a different differentiation variable which affects the
evaluation rules, and similarly in a restriction propagation algorithm,
the restricted type provides which side the terminals of the subexpression
should be restricted to. We therefore partition $T_O$
further into algorithm-specific disjoint sets $T_V$ and $T_C$, where
$T_V$ includes types of expressions that can be evaluated given the
values of its operands as in~\eqref{eq:recursiveevaluateoperator}, and
$T_C$ includes types of expressions which provide a context
for the evaluation of the operands. By defining corresponding
evaluation operators $E_V$ and $E_C$, we can then
extend~\eqref{eq:recursiveevaluateoperator} to
\begin{align}
  \label{eq:flexibleevaluateoperator}
  \tilde{E}_R(e_i) &= \begin{cases}
    E_T(e_i),
    \qquad & \text{if} \, \type(e_i) \in T_T, \\
    E_V\left(e_i, \langle \tilde{E}_R(e_{I^i_j}) \rangle_{j=1}^{p_i}\right),
    \qquad & \text{if} \, \type(e_i) \in T_V, \\
    E_C\left(e_i, \langle e_{I^i_j} \rangle_{j=1}^{p_i}\right),
    \qquad & \text{if} \, \type(e_i) \in T_C.
  \end{cases}
\end{align}
Note that the evaluation operator $E_C$ may make further recursive
calls to $\tilde{E}_R$ or a related recursive algorithm, but it is
assumed that its operands cannot be pre-evaluated without the provided
context. The difference between how $E_V$ and $E_C$ are applied
corresponds to a post-order versus pre-order evaluation of the DAG
vertices.

The recursive operator $E_R$ defined
by~\eqref{eq:recursiveevaluateoperator} can be implemented by
traversing the link-based DAG representation with a post-order
traversal algorithm, to recursively visit and evaluate child vertices
before their parent. The more flexible operator $\tilde{E}_R$ defined
by~\eqref{eq:flexibleevaluateoperator} can be implemented similarly
but with a mix of post-order and pre-order traversals depending on the
visited types. The evaluation of $e$ can then be written simply
\begin{align}
  \label{eq:evaluaterecursive}
  v = \tilde{E}_R(e) \equiv \tilde{E}_R(e_{m+n}).
\end{align}

The simpler recursive operator~\eqref{eq:recursiveevaluateoperator}
can also be implemented as a loop over subexpressions in a topological
ordering, as shown in Algorithm~\ref{alg:evaluate}.  This
non-recursive algorithm can be implemented by first constructing the
list-based DAG representation and then iterating over the vertices.
Using the list-based representation and the non-recursive algorithm
has the advantage of never visiting a vertex twice, even if it is
reachable through multiple paths. We remark that a cache mechanism
may, however, remove such duplicate evaluation in the recursive
implementation as well. The list-based representation also assigns an
integer label to each vertex which can be used to store associated
data efficiently in arrays during the algorithm.  However, the
construction of the list-based DAG representation is not free, and the
fixed labeling is a downside in algorithms where new expressions are
constructed. In the following exposition, the choice of algorithm
structure, recursive or non-recursive, is considered mainly an
implementation detail, controlled by performance and convenience
considerations and not by functionality.

\begin{algorithm}
  \begin{algorithmic}[1]
    \For{$i = 1, \ldots, m$}
    \State $v_i = E_T(e_i)$
    \EndFor
    \For{$i = m+1, \ldots, m+n$}
    \State $v_i = E_O(e_i, \langle v_{I^i_j} \rangle_{j=1}^{p_i})$
    \EndFor
    \State{$v := v_{m+n}$}
  \end{algorithmic}
  \caption{Non-recursive algorithm for evaluation of expressions.}
  \label{alg:evaluate}
\end{algorithm}

\subsection{Type based function dispatch and the visitor pattern}

In an implementation of the evaluation algorithm described in the
previous section, the specific evaluation actions must be selected
dynamically based on the type of the expression argument. By
subclassing the UFL provided class \uflc{MultiFunction} and
implementing a handler function for each expression type, calls to an
instance of this class are dynamically dispatched to the correct
handler based on the type of the first argument. If a handler is
missing, the closest superclass handler is used instead, which makes
it easy to implement default rules for groups of types. An example is
shown below.

\inputpython{snippets/multifunction.py}

Building on this same dynamic multifunction design and the Visitor
pattern~\citep{GammaEtAl1993}, the class \uflc{Transformer} can be
subclassed in the same way to implement many recursive symbolic
algorithms following the structure
of~\eqref{eq:recursiveevaluateoperator}
or~\eqref{eq:flexibleevaluateoperator}.  Calling upon an object of a
\uflc{Transformer} subclass to visit an expression will result in a
recursive application of type-specific rules to subexpressions.  The
example below shows a numerical evaluation of a simple expression,
using a pure post-order implementation as
in~\eqref{eq:recursiveevaluateoperator}. Whether to visit expressions
post- or pre-order is specified per handler simply by taking visited
expressions for child nodes as arguments or not. In more detail, and
with reference to the example below, a type will be placed in the
$T_C$ set if the corresponding handler omits the \uflc{*values}
argument. The \uflc{visit} method will then automatically call the
handler without first handling the operands of its argument.

\inputpython{snippets/transformer.py}

\subsection{Partial evaluation}
\label{sec:partialevaluation}

Some symbolic algorithms involve modification of subexpressions, and
such algorithms share a need to apply an operator to a new sequence of
operands. We will designate the notation $\type(e)(\langle f_j
\rangle_j)$ to the construction of an operator of the same type as $e$
with the given operands.  If the operands are unchanged, the original
expression can be reused since all expressions are considered
immutable, thus saving memory. This can easily be accomplished for
UFL-based algorithms by subclassing the \uflc{ReuseTransformer} class.
In this case, the algorithm inherits the fallback rules given by
\begin{align}
  \label{eq:reusetransformer}
  E_{\mathrm{reuse}}\left(e_i\right)
  &= \begin{cases}
    e_i,
    \qquad & \text{if} \, \type(e_i) \in T_T, \\
    \type(e_i)\left(\langle E_{\mathrm{self}}(e_{I_j^i}) \rangle_{j=1}^{p_i}\right),
    \qquad & \text{if} \, \type(e_i) \in T_O,
  \end{cases}
\end{align}
where $E_{\mathrm{self}}$ refers to $E_{\mathrm{reuse}}$ or overridden rules in a
subclass.  Based on $E_{\mathrm{reuse}}$, algorithms can be written to just
modify what they need and let the fallback rules
in~\eqref{eq:reusetransformer} rebuild the surrounding expression with
no additional algorithm-specific code.  This allows a very compact
implementation for algorithms such as the partial evaluation in which
terminal expressions are replaced with other expressions through a
given mapping. As an example, consider a partial evaluation algorithm
mathematically described by:
\begin{align}
  \label{eq:replacer}
  E_{\mathrm{replace}}\left(e_i\right)
  &= \begin{cases}
    \operatorname{map}(e_i),
    \qquad & \text{if} \, \type(e_i) \in T_T, \\
    \text{delegate to} \, E_{\mathrm{reuse}}(e_i),
    \qquad & \text{otherwise},
  \end{cases}
\end{align}
where \emph{delegate to $E_{\mathrm{reuse}}(e_i)$} represents
delegation to inherited rules from the superclass. This
can be compactly implemented in UFL as follows:
\inputpython{snippets/reusetransformer.py}

\subsection{Differentiation}
\label{sec:derivatives}

The differentiation algorithm in UFL is a two-level algorithm. The
two levels are used to handle expressions involving derivatives with
respect to different types of variables. In a first step, a simple outer
algorithm is employed to evaluate the innermost derivatives first; that
is, the derivative expressions closest to the terminal expressions. This
outer algorithm then calls a single-variable differentiation algorithm
for each derivative expression visited. This, in turn, allows the inner
algorithm to assume that no nested derivatives are encountered. Thus,
for instance in the evaluation of $d(c \Grad(v u)) / dv$,
an inner algorithm is called first to evaluate $\Grad(v u)$, and second
to evaluate $d(c (\Grad(v) u + v \Grad(u))) / dv$. Note
that the $\Grad$ operator is kept in the expression DAG after derivative
evaluation, but is guaranteed to only apply directly to spatially
varying terminal expressions.  More sophisticated approaches to
nested differentiation have been explored in \citet{Karczmarczuk2001},
\citet{PearlmutterSiskind2007} and \citet{SiskindPearlmutter2008},
however we have considered this additional complexity unnecessary for
the purpose of UFL.

The inner algorithm handles differentiation of an expression $e$ with
respect to a single differentiation variable~$u$. Both $e$ and $u$ may
be tensor-valued, but in the following we illustrate using scalars for
the sake of clarity. By setting the value in
Algorithm~\ref{alg:evaluate} to a tuple of the subexpression and its
derivative: $v_i = (e_i, de_i/du)$, we obtain the standard Forward-mode
Automatic Differentiation algorithm\footnote{However, output of
  the algorithm is a new symbolic UFL expression, and the algorithm is
  therefore clearly a symbolic differentiation algorithm. In UFL
  context, the distinction between Automatic Differentiation and
  Symbolic Differentiation is therefore irrelevant.}
(see~\citep{Griewank1989}).
This algorithm can also be written in recursive form as
in~\eqref{eq:recursiveevaluateoperator}.  However, because of a few
exceptions specific to differentiation variable types, the actual
algorithm in UFL requires the more flexible framework given
by~\eqref{eq:flexibleevaluateoperator}. For simplicity, we will use
$v_i = de_i/du$ to define the evaluation rules below.

Generic differentiation rules are implemented as handler functions in
a transformer class corresponding to
\begin{align}
E_{\mathrm{AD}}(e_i) & =
\begin{cases}
0,
  \qquad & \text{if} \ \type(e_i) \in T_T, \\
\sum_{j=1}^{p_i} \frac{\partial e_i}{\partial e_{I_j^i}} E_{\mathrm{AD}}(e_{I_j^i}),
  \qquad & \text{if} \ \type(e_i) \in T_O.
\end{cases}
\end{align}
For each type of differentiation variable, the default differentiation
rules in $E_{\mathrm{AD}}$ are subclassed to encode the dependency of
expression types with respect to the differentiation variable.

For spatial derivatives, the full gradient is used to represent the
derivatives of functions, giving the evaluation operator:
\begin{align}
E_{\mathrm{XD}}(e) & =
\begin{cases}
 I,             \qquad & \text{if }e\text{ is the spatial coordinate vector}, \\
 0,             \qquad & \text{if }e\text{ is a piecewise constant function}, \\
 \Grad e,       \qquad & \text{if }e\text{ is a non-constant function}, \\
 \Grad \Grad f, \qquad & \text{if }e\text{ is} \, \Grad f, \\
\text{delegate to} \, E_{\mathrm{AD}}(e),           \qquad & \text{otherwise}.
\end{cases}
\end{align}

For differentiation with respect to user-annotated variables
(see~\ref{subsec:diffopandvariables}), the operator rules are instead
modified as:
\begin{align}
E_{\mathrm{VD}}(e) & =
  \begin{cases}
   1,                        \qquad & \text{if }e\text{ is the variable instance} \, u, \\
   E_{\mathrm{VD}}(f),       \qquad & \text{if }e\text{ \small is another variable instance annotating the expression } f, \\
  \text{delegate to} \, E_{\mathrm{AD}}(e),                    \qquad & \text{otherwise}.
  \end{cases}
\end{align}
Note that in this case the \uflc{Variable} type lies in the $T_C$
set, and is thus visited before the annotated expression. Hence, the
underlying expression is visited by a further recursive call to
$E_{\mathrm{VD}}(f)$ only if the variable $e$ is different from~$u$.

Finally, for directional derivatives of an expression $e$ with respect
to a coefficient function $u$ in the direction of $v$, with possibly
user-specified $\partial g / \partial u = h$ for a subexpression
$g$, the rules become:
\begin{align}
  E_{\mathrm{DD}}(e) & =
  \begin{cases}
    v,             \qquad & \text{if }e\text{ is the function} \, u, \\
    \Grad E_{\mathrm{DD}}(f),  \qquad & \text{if }e\text{ is} \, \Grad f, \\
    h v,           \qquad & \text{if }e\text{ is the function} \, g, \\
    \text{delegate to} \, E_{\mathrm{AD}}(e),           \qquad & \text{otherwise}.
  \end{cases}
\end{align}
Note that in this case, we consider the gradient of a function as a
terminal entity, although it is represented as two expression nodes.

To support differentiation with respect to specific components of a
mixed function (or a vector-valued function), the same rules can be
applied by choosing an appropriate expression for~$v$. As an example,
consider the functions
\begin{align}
  u: X \rightarrow \R^3, \qquad \hat{u}=(u_1, u_3), \qquad \hat{v}: X \rightarrow \R^2.
\end{align}
To compute $D_{\hat{u}} M(u) [\hat{v}]$, the differentiation rule for
$u$ must yield a vector-valued derivative of the same value shape as
$u$. This is accomplished by padding $\hat{v}$ and using
$v=(\hat{v}_1, 0, \hat{v}_2)$ as the direction. That is,
\begin{align}
D_{\hat{u}} M(u) \left[\hat{v}\right]
  = \frac{d}{d\tau} \left[
     M((u_1+\tau \hat{v}_1, u_2+0 \tau, u_3+\tau \hat{v}_2))
     \right]_{\tau=0}
  = D_{u} M(u) \left[v\right].
\end{align}
This concept of padding to support component-wise derivatives extends
to arbitrary tensor-valued functions and functions in mixed element
spaces, as well as differentiation of variable components that are
part of different functions.

We conclude this section by commenting on the relation between the UFL
differentiation algorithms and the algorithms implemented in
Sundance~\citep{long:2010}. In the Sundance approach, the derivative
is computed numerically on the fly, and the use of BLAS amortizes the
cost of expression DAG traversal at run time. However, the traversal
cost does increase with the size of the expression DAG. Therefore,
Sundance avoids computing the DAG for the
derivative. In contrast, as UFL is typically used in combination with
code generation tools, we may differentiate and then simplify. This
allows us to produce the expression DAG for the derivative and then
produce efficient code for it. We can see the symbolic traversal cost
as a part of the software build time, and it does not affect the
runtime for computing/assembling variational forms.

\section{Validation}
\label{sec:validation}

Several validation steps are performed by UFL at the stage where an
operator is applied to an expression. All UFL operator types validate
the properties of its operands in various ways to ensure that the
expressions are meaningful. Most importantly, each operator validates
the operand value shapes and verifies that the use of free-indices is
consistent. This type of validation catches common indexing bugs at an
early stage. Other examples of validations include: checking that
value shapes and indices match when adding expressions; checking value
shapes for tensor algebra products;
and checking of index ranges for explicit indexing of tensor-valued expressions.
Most indexing in UFL
expressions uses implicit ranges, which reduces repetition and common
sources of errors. When defining a form, an integrand expression must
always be scalar-valued without any free-indices. This is also
checked.  When the form is preprocessed, prior to form compilation, a
number of properties are checked such as: all integrals must depend
linearly on the same set of argument functions; and in interior
facet integrals all functions must be restricted. The latter in
particular forces increased clarity in the formulations.

Testing of symbolic frameworks is hard because every
algorithm must be tested with an appropriate selection of expression
type combinations to achieve high test coverage. In an attempt to
answer to this challenge in UFL, we have used multiple layers of
defensive programming with assertions, unit testing and integration
testing in other FEniCS components. Static code analysis with
PyChecker~\citep{pychecker:www} was very useful during the main
development phase. Algorithms in UFL are sprinkled with assertions to
document assumptions and catch any that fail. Unit tests cover many
common (and uncommon) combinations of operators and applications of
algorithms such as differentiation. In addition to the unit tests in
UFL, unit and regression tests in FFC and DOLFIN test the use of UFL
for many PDE application examples.

As an example for testing differentiation, a numerical evaluation of
the symbolic derivative can be compared with the evaluation of
manually derived derivative expressions or reference values. Of
particular interest when considering validation of UFL is a set of
integration tests in DOLFIN which exploit Green's theorem in 1D or $n$D
for a scalar function $f(x)$ or a vector-valued function $v(x)$, that
is
\begin{equation}
  \int_a^b f^\prime(x) \dx = \left[f(x)\right]_a^b, \quad
  \int_\Omega \Div v \dx = \int_{\partial\Omega} v \cdot n \ds.
\label{eq:greens}
\end{equation}
This identity is ideal for combined testing of symbolic
differentiation, numerical differentiation, and symbolic evaluation,
or even higher order derivatives by setting $v = \Grad f$
in~\eqref{eq:greens}. Exploiting such mathematical identities is key
to robust testing of mathematical software, and combining the symbolic
and numerical paradigms provides good opportunities for discovering
errors.

A software stack such as that provided in the FEniCS Project with (i)
a DSL and symbolic framework in UFL, (ii) automated code generation in
the form compiler FFC~\citep{logg:2012}, and (iii) library code in the
problem solving environment DOLFIN~\citep{logg:2012b}, naturally
introduces additional complexity to the debugging process if something
goes wrong. However, the high abstraction level allows us to check the
correctness of end-user programs in various ways. Automated validation
of expressions and forms by UFL allows consistency checks and catching
of user errors at various levels of abstraction.  Last, but not least,
UFL has been tested in active use by researchers for more than three
years as part of the FEniCS Project.

\section{Conclusions}
\label{sec:conclusions}

We have presented the Unified Form Language and shown how the language
and its associated algorithms allow compact, readable and efficient
specification of mathematical expressions related to variational
formulations of partial differential equations. We have presented both
high-level and detailed views of UFL to communicate its practical use
and to provide developers and technical users a firm grounding in the
design principles of UFL for understanding and building upon UFL.

UFL is a stand-alone Python module that has been extensively used as part
of the FEniCS software pipeline since 2009. The UFL functionality has been
crucial in enabling advanced automated finite element algorithms in the
FEniCS context, especially for complicated coupled systems of equations
and for problems for which automatic differentiation dramatically
reduced the burden on the application developer.  UFL has also proven
to be extensible beyond the core implementation, as exemplified by
\citet{NikbakhtWells2009} and \citet{MassingEtAl2012} in the context
of extended finite element methods, and \citet{MarkallEtAl2012} in
relation to code generation for different architectures.  UFL is an
actively developed project and continues to further extend the power
and expressiveness of the language.

\begin{acks}
The authors wish to thank Kent-Andre Mardal and Johannes Ring for
their contributions to UFL, and Pearu Peterson for discussions about
symbolic representations during the initial design phase.
\end{acks}
\bibliographystyle{abbrvnat}
\bibliography{references}
\end{document}